\documentclass[a4paper,english,aps,floats,onecolumn,showpacs,nofootinbib,preprintnumbers]{revtex4}
\usepackage{graphicx}
\usepackage{subfig}
\usepackage[breaklinks,linktoc=page]{hyperref}

\newcommand{\nc}{\newcommand}
\nc{\bea}{\begin{eqnarray}}
\nc{\eea}{\end{eqnarray}}
\nc{\be}[1]{\begin{equation} \mbox{$\label{#1}$}}
\nc{\ee}{\vspace{0.1cm}\end{equation}}
\nc{\eq}[1]{\mbox{Eq.\ (\ref{#1})}}
\nc{\fig}[1]{\mbox{Fig.\ (\ref{#1})}}
\nc{\ch}[1]{\mbox{Chapter\ \ref{#1}}}
\nc{\sect}[1]{\mbox{Section\ \ref{#1}}}
\nc{\HRule}{\rule{\linewidth}{0.5mm}}

\nc{\lh}{\lambda_h}
\nc{\ls}{\lambda_s}
\nc{\lhs}{\lambda_{hs}}
\nc{\lp}{\lambda_\phi}
\nc{\tdeps}{\tilde{\epsilon}}
\nc{\tdeta}{\tilde{\eta}}
\nc{\tdep}{\tilde{\epsilon}}
\nc{\tdu}{\tilde{U}}
\nc{\tdN}{\tilde{N}}
\nc{\tda}{\tilde{a}}
\nc{\tdH}{\tilde{H}}
\nc{\tdrho}{\tilde{\rho}}
\nc{\tdze}{\tilde{\zeta^2}}
\nc{\tdpartu}{\tilde{\partial_{\mu}}}
\nc{\tdpartd}{\tilde{\partial^{\mu}}}
\def\mso{m_{s_{o}}}
\nc{\cv}{{\cal P}_{\zeta}(k)}

\nc{\Aikp}{{\rm Ai}'(-\kappa^2)}
\nc{\Aik}{{\rm Ai}(-\kappa^2)}
\nc{\Bikp}{{\rm Bi}'(-\kappa^2)}
\nc{\Bik}{{\rm Bi}(-\kappa^2)}

\nc{\hf}{\frac{1}{2}}
\nc{\hfrt}{\frac{1}{\sqrt{2}}}

\def\GeV{{\rm \ GeV}}

\def\mso{m_{s_{o}}}

\def\bibnamefont#1{#1}
\def\bibfnamefont#1{#1}
\def\citenamefont#1{#1}
\def\url#1{\texttt{#1}}

\begin{document}
\title{Distinguishing Higgs inflation and its variants}

\author{Rose N. Lerner}
\email{rose.lerner@helsinki.fi}
\affiliation{Physics Department, University of Helsinki, FIN-00014, Finland}
\affiliation{Helsinki Institute of Physics, University of Helsinki, FIN-00014, Finland}

\author{John McDonald}
\email{j.mcdonald@lancaster.ac.uk}
\affiliation{Cosmology and Astroparticle Physics Group, University of Lancaster, Lancaster LA1 4YB, UK}

\preprint{HIP-2011-10/TH}

\begin{abstract}

We consider how Higgs Inflation can be observationally distinguished from variants based on gauge singlet scalar extensions of the Standard Model, in particular where the inflaton is a non-minimally coupled gauge singlet scalar ($S$-inflation). We show that radiative corrections generally cause the spectral index $n$ to decrease relative to the classical value as the Higgs mass is increased if the Higgs boson is the inflaton, whereas $n$ increases with increasing Higgs mass if the inflaton is a gauge singlet scalar. The accuracy to which $n$ can be calculated in these models depends on how precisely the reheating temperature can be determined. The number of Einstein frame e-foldings $\tdN$ is similar in both models, with $\tdN \approx 58-61$ for singlet inflation compared with $\tdN \approx 57-60$ for Higgs inflation. This allows the spectral index to be calculated to an accuracy $\Delta n = \pm 0.001$.
 Provided the Higgs mass is above $\approx 135\GeV$, a combination of a Higgs mass measurement and a precise determination of $n$ will enable
Higgs Inflation and $S$-inflation to be distinguished.

\end{abstract}

\maketitle

\section{Introduction}

   At present our knowledge of particle physics is entirely summarised by the Standard Model (SM). There is no direct evidence of particle physics at mass scales greater than the weak scale, although the SM leaves many unanswered questions, such as the origin of the baryon asymmetry, the nature of dark matter, the solution to the strong CP problem, the detailed structure of gauge and Yukawa coupling constants, the origin of neutrino masses and a particle-physics interpretation of inflation in the early universe. Nevertheless, since the only particle physics mass scale we know to exist is the weak scale, it makes sense to explore the possibility that the answer to these questions can be obtained from models of weak scale particle physics. This has the added appeal that such models may be testable at colliders.

         Two of these questions might be solved by scalar fields with masses and interactions characterised by the weak scale. Firstly, Higgs Inflation \cite{Bezrukov:2007ep} proposes that the SM Higgs boson could be responsible for inflation. This is possible if there is a large dimensionless coupling of the Higgs to the Ricci scalar through the term $\xi_h H^{\dagger}H R$, with $\xi_h \sim 10^4$. Secondly, a real gauge singlet scalar interacting with the SM via the `Higgs portal' provides a minimal extension of the SM with a thermal relic WIMP dark matter candidate \cite{McDonald:1993ex,Silveira:1985rk,McDonald:1993ey,Burgess:2000yq,Bento:2001yk}. In this case there is a second possibility for inflation, since the singlet scalar could also have a non-minimal coupling to gravity and so provide an alternative to Higgs Inflation, which we call $S$-inflation \cite{Lerner:2009xg}. Higgs Inflation in a gauge singlet extension of the Standard Model has been discussed in \cite{Clark:2009dc}.

          In all of these models the scalar couplings are ${\cal O}(0.1)$ and the potential is made flat by choosing a large non-minimal coupling $\xi_\phi$ of the inflaton\footnote{
In the following, $\phi$ is used to refer to the inflaton in general, which can be either $s$ or $h$, depending which version of the model we are considering.} $\phi$ to gravity, such that
\be{COBEnorm}
 \frac{\lambda_{\phi}}{\xi_{\phi}^2} \simeq \frac{3(0.027)^4}{\tdN^2}
\ee
where $\lambda_\phi$ is the quartic self-coupling of the inflaton and $\tdN$ is the number of Einstein frame e-foldings between the WMAP pivot scale leaving the horizon and the end of inflation. This large non-minimal coupling to gravity causes the theory to be effectively scale-invariant at large values of $\phi \approx {\cal O}(0.1)M_p$. Thus slow roll inflation can take place.

     It is remarkable that inflation models based on weak scale particle
theories have the predictive power to connect observables from cosmology (the spectral index) and particle physics (the Higgs mass). It is therefore important to clarify the predictions
for the spectral index as a function of the Higgs mass $m_h$, as the models discussed in this paper will be within the reach of Planck and the LHC in the next few years.

    At the classical level, Higgs Inflation and its singlet variants make the same predictions for the inflation observables. At the quantum level, the spectral index of the models as a function of Higgs mass will differ. The predictions for Higgs inflation have been discussed in \cite{Bezrukov:2008ut,Bezrukov:2008ej,Bezrukov:2009db,Barvinsky:2008ia,Barvinsky:2009fy,DeSimone:2008ei}.
Our goal in this paper is to calculate the predictions for the spectral index and its dependence on the Higgs mass in the different models using the same method throughout.  In this way we can be sure that any differences in the predictions are due to the fundamental differences between the models and not due to the initial conditions, method of obtaining the effective potential or specifics of the code used. We will show that a combination of Higgs mass measurement and a precise determination of $n$ can distinguish between $S$-inflation and models where the inflaton is the Higgs boson.

    It has been suggested that Higgs Inflation and its singlet variants may not be consistent particle physics theories at the quantum level (see \cite{Barbon:2009ya,Burgess:2010zq,Lerner:2009na,Hertzberg:2010dc,Kaiser:2010ps,Bezrukov:2010jz} and references within for discussion). It is known that perturbative unitarity is violated in tree-level graviton exchange-mediated scalar particle scattering via the non-minimal coupling to gravity\footnote{It has also been suggested that the non-polynomial potential in the Einstein frame could also lead to unitarity violation at $E \approx \Lambda$. We believe that this will not be the case if the non-polynomial potential of the inflaton interpolates between renormalizable polynomial potentials, since we would expect scattering processes at $E \gg \Lambda$ be dominated by the renormalizable potential at large $\phi \sim E$  up to small corrections proportional to powers of $\Lambda/E$, vanishing as $\Lambda \rightarrow 0$.} at $E \approx \Lambda = \frac{M_{p}}{\xi_{\phi}}$ \cite{Hertzberg:2010dc}. This scale $\Lambda$ is below the scale of inflation. Therefore, something must happen at the scale $\Lambda$ --- either  strong coupling ensures that {\em true} non-perturbative unitarity is not violated even though it is at tree-level, or some UV completion of the theory must restore unitarity. If new physics is necessary, then the model is no longer natural as we would need to add terms to the Lagrangian which would be large at the scale of inflation. However, if the theory is non-perturbative with no new physics, then we {\em can} make predictions from inflation, as this is not directly related to particle scattering processes. The possibility of strong coupling as a solution to unitarity has been discussed in \cite{Bezrukov:2009db} and
\cite{hw}\footnote{An argument against this point of view is presented in \cite{Atkins:2010yg,Atkins:2010re}.}.

Therefore, if strong coupling unitarizes the theory, Higgs Inflation and its variants would be consistent theories, requiring no new particles or interactions\footnote{However, if unitarity is really violated in scalar scattering processes, it may still be possible to add interactions and particles to unitarize the theory \cite{Giudice:2010ka,Lerner:2010mq}.}. This can be tested by the resulting predictions of the model. Our philosophy
for the remainder of this paper is to consider the possibility that Higgs inflation and its variants are consistent theories and therefore should be subject to rigourous testing against experimental data.

Our paper is organised as follows. In Section 2 we introduce the three models of non-minimally coupled inflation that we consider:  (i) original Higgs Inflation with no additional particles beyond the Standard Model  \cite{Bezrukov:2007ep}, (ii) inflation along the Higgs direction in an extension with a gauge singlet scalar  \cite{Clark:2009dc} and (iii) $S$-inflation \cite{Lerner:2009xg}. In Section 3 we compute the quantum effective potential for the models.  In Section 4 we discuss reheating. This is crucial, as the ability to accurately predict the reheating temperature is a key feature of this class of models, allowing the
number of e-foldings of inflation for a given length scale to be precisely determined and so the spectral index to be precisely predicted.  In Section 5 we compare the predictions of Higgs Inflation and $S$-inflation at the quantum level.
In Section 6 we present our conclusions.

\section{Non-minimally coupled models of inflation}
\label{SEC:nonmin}

\subsection*{Higgs inflation action in the Jordan frame}

  We consider three variants of the Higgs inflation model, which can all be described by the following action for different choices of the couplings and inflaton field. The Jordan frame action, including quantum corrections, is
\bea
\label{Jaction}
   S_J =  \int \sqrt{-\!g}\,d^4\! x \Big({\cal L}_{\overline{SM}} - \frac{M_p^2R}{2}  - \xi_h G_H^2 H^{\dagger}H R - \hf\xi_s s^2R
+ G_H^2\left(D_\mu H\right)^{\dagger}\left(D^\mu H\right) + \hf\left(\partial_\mu s\right)\left(\partial^\mu s\right) - V(H^{\dagger}H,s^2) \Big),    ~\eea
where ${\cal L}_{\overline{SM}}$ is the Standard Model Lagrangian density minus the purely Higgs doublet terms and $V(H^{\dagger}H,s^2)$ is the renormalisation group (RG) improved effective potential, given by
\bea
\label{Jpot}
V  =  \lh \left(G_H^2\left(H^{\dagger}H\right) - \frac{v^2}{2}\right)^2   + \frac{\lhs}{2} G_H^2 s^2 H^{\dagger}H
+ \frac{\ls s^4}{4}  + \hf \mso^2 s^2.  ~\eea
$\lhs$ is the Higgs portal coupling.
$ G_H(t) = \exp{\left( -\int_0^t{ \frac{dt'\gamma_h(t')}{1+\gamma_h(t')} }\right)}$ and the anomalous dimension $\gamma_h$ is given by \eq{EQ:gamma}, while $G_s = 1$ (as $\gamma_s = 0$). For `pure' Higgs inflation, we set $s = \ls=\lhs=\xi_s = 0$. We also consider the case where the inflaton is the Higgs boson in the presence of a singlet scalar \cite{Clark:2009dc}, setting $\xi_s(m_t) = 0$,  and the case where the singlet is the inflaton \cite{Lerner:2009xg}, setting $\xi_h(m_t) = 0$. Setting the non-inflationary non-minimal coupling to zero is a reasonable simplification. If $\xi_s$ and $\xi_h$ are of similar magnitudes, then the potential in the Einstein frame ($\propto \xi^{-1}_{\phi}$) would be similar in all directions and inflation would be expected to occur in a general direction in the $s-h$ plane. Our aim is to compare the two limiting cases where $h$ and $s$ are the inflaton respectively, so fixing the non-inflationary $\xi_\phi$ to be small (and so the potential in that direction to be large),  ensures that inflation will occur either along the Higgs or the $s$ direction.

\subsection*{Conformal transformation to the Einstein frame}

We will transform the entire RG improved action to the Einstein frame, where the fields are minimally coupled to gravity. This will allow us to use the standard slow roll formalism to calculate the spectral index $n$ and tensor-to-scalar ratio $r$. Quantities in the Einstein frame will be denoted by a tilde (e.g. $\tilde{g}_{\mu\nu}$).
For now we consider only the physical Higgs field $h$, where $H~=~\frac{1}{\sqrt{2}} \left(\begin{array}{c} 0 \\ h + v \end{array}\right) $ and $h$ is real. The inclusion of the non-physical components of the Higgs field is important when computing quantum corrections to the effective potential and also in the analysis of unitarity-conservation, but they are not important for the dynamics of the inflaton field and the calculation of the spectral index.

    The Jordan frame and the Einstein frame are related by a conformal transformation which transforms the metric in a field dependent way. Considering the action \eq{Jaction}, for general $h$ and $s$, the conformal transformation to the Einstein frame is defined by
\be{2} \tilde{g}_{\mu\nu} = \Omega ^2 g_{\mu\nu} \ee with
\be{3} \Omega^2 = 1 + \frac{\xi_h G_H^2 h^2}{M_p^2} + \frac{\xi_s s^2}{M_p^2}.\ee
Then
\bea \label{e4}
S  &=  &\int d^4x\sqrt{-\tilde{g}}  \left[  \tilde{{\cal L}}_{\overline{SM}} + \frac{1}{2}\left(\frac{1}{\Omega^2} + \frac{6\xi_s^2s^2}{M_p^2\Omega^4}\right)\tilde{g}^{\mu\nu}\partial_\mu s \partial_\nu s  + \frac{1}{2}\left(\frac{G_H^2\Omega ^2 + 6 M_p^2 \Omega^2\left(\frac{d\Omega}{d h}\right)^2}{\Omega ^4}\right) \tilde{g}^{\mu\nu}\partial_\mu h \partial_\nu h \right.\nonumber \\
& &    {} + \frac{6\xi_s\xi_h G_H^2s\;h \; \tilde{g}^{\mu\nu}\;\partial_\mu s\partial_\nu h}{M_p^2\Omega^4} \left. -\frac{M_P^2\tilde{R}}{2}- \frac{V(s,h)}{\Omega^4}\right]
\eea
where $\tilde{R}$ is the Ricci scalar with respect to $\tilde{g}_{\mu \nu}$ and
\be{imp5}
\frac{d\Omega}{d h} = \frac{1}{2\Omega} \frac{\xi_h h G_H^2}{M_p^2}\left( 2 - \frac{2\gamma_H}{1+\gamma_H} + \frac{1}{\xi_h}\frac{d\xi_h}{dt}\right).
\ee We can then rescale the fields using{\footnote{These are only total derivatives in the limits $s\rightarrow 0$ and $h \rightarrow 0$ respectively.}}
\be{4}
\frac{d\chi_h}{dh} = \sqrt{\frac{G_H^2\Omega ^2 + 6 M_p^2 \Omega^2\left(\frac{d\Omega}{d h}\right)^2}{\Omega ^4}} \;\; ; \;\;\; \frac{d\chi_s}{ds} = \sqrt{\frac{\Omega ^2 + 6 \xi_s ^2s^2/M_P^2}{\Omega ^4}}
~\ee
to give
\bea
S_E =  \int d^4x\sqrt{-\tilde{g}}\Big( \tilde{{\cal L}}_{\overline{SM}} - \frac{M_P^2\tilde{R}}{2} + \frac{1}{2}\tilde{g}^{\mu\nu} \partial_\mu \chi_h \partial_\nu \chi_h + \frac{1}{2}\tilde{g}^{\mu\nu}\partial _\mu \chi_s \partial_\nu \chi_s
+ A(\chi_s,\chi_h)\tilde{g}^{\mu\nu}\partial_\mu \chi_h \partial_\nu \chi_s - U(\chi_s,\chi_h)\Big),
~\eea
where
\be{6} A(\chi_s,\chi_h) = \frac{6G_H^2\xi_s\xi_h}{M_P^2 \Omega^4}\frac{ds}{d\chi_s} \frac{dh}{d\chi_h} h s  \ee
and
\be{poteq} U(\chi_s,\chi_h) \simeq \frac{1}{\Omega^4} \left( \frac{\lambda_h}{4}G_H^4 h^4  + \frac{\lambda_s}{4}^4 s^4 + \frac{\lhs}{4} G_H^2 s^2 h^2 \right)
~\ee

During inflation, only one field (the inflaton) is non-zero and so the non-canonically normalised term $A$ is zero. After inflation, the Jordan and Einstein frames will be indistinguishable (as both $\frac{H^\dagger H}{\xi_h} \ll M_p$ and $\frac{s^2}{\xi_s} \ll M_p$). Therefore the curvature perturbation spectrum calculated in the Einstein frame can be compared to measurements made in the physical Jordan frame.

\subsection*{Classical predictions from slow-roll inflation}

Using the tree-level potential and the approximation $\frac{\xi_\phi \phi^2}{M_p^2}~\gg~1$, the tree-level slow-roll parameters are
\be{c1} \tdeps \simeq \frac{4}{3}\frac{M_p^4}{\xi_\phi^2 \phi^4} \;\; ; \;\;\;  \tdeta \simeq -\frac{4}{3}\frac{M_p^2}{\xi_\phi \phi^2}  ~,\ee where $\phi_{\tilde{N}}^2 \approx 4 M_P^2 \tilde{N}/3 \xi_\phi$ is the field at $\tdN$ e-foldings. A calculation of the classical spectral index and tensor-to-scalar ratio then gives
\bea
\label{EQ:SRtree}
n_{cl} &\approx& 1 -\frac{2}{\tdN} - \frac{3}{2 \tdN^2} + {\cal O}\left(\frac{1}{\tdN^3}\right) = 0.965 ~;\\
r &\approx &\frac{12}{\tdN^2}  + {\cal O}\left(\frac{1}{\xi_s \tdN^2}\right) = 3.6\times 10^{-3}  , \eea
where we have used $\tdN = 58$. This is close to the WMAP pivot scale in these models. The tree-level potential and classical predictions are identical for all non-minimally coupled scalar field models with a $\phi^2 R$ coupling to gravity and $\phi^4$ potential at large $\phi$.

\section{Quantum corrections to the scalar potential}
\label{radcor}

In this section we compute the RG improved effective potential for Higgs Inflation and $S$-inflation. We calculate the effective potential in the Jordan frame using the Feynman rules for the SM fields defined in the Jordan frame. In this approach, the effect of the non-minimal coupling to gravity is taken into account by using a modified propagator for the non-minimally coupled inflaton field. This is derived by quantising the inflaton field in the Einstein frame but without rescaling the field to the canonically normalised form. (We review this in Appendix A.) Using these Feynman rules, the RG equations for the couplings are derived and the
effective potential is calculated in the Jordan frame. This is then transformed to the Einstein frame, where it is used to study slow-roll inflation. While it is also possible to use the 1-loop Coleman-Weinberg potential, this would not be the standard flat-space form but must also include the contribution of the non-minimal coupling\footnote{In our previous paper on $S$-inflation, the contribution of the non-minimal coupling to the Coleman-Weinberg potential was not included, but this has a small effect in the case of a real gauge singlet scalar inflaton.} of $\phi$ to $R$.

\subsection*{Initial conditions}

We choose $\xi_{\phi}(m_t)$ such that the model is correctly normalised to the WMAP 7-year mean value for the curvature perturbation $\Delta^2_{\cal R}$
\cite{Komatsu:2010fb,Lyth:1998xn}, which requires
\be{EQ:newWMAP}
\frac{U}{\tdeps} = (0.00275M_p)^4 .
\ee
The initial values of the coupling constants are defined at the renormalisation scale $\mu = m_t$, with $m_t = 171.0$ GeV and $v = 246.22$ GeV. The gauge couplings are given by
\be{ICgauge} \frac{g^2(m_t)}{4\pi} = 0.03344,\;\;\;  \frac{g'^2(m_t)}{4\pi} = 0.01027 \;\;\; ~\mbox{and}~\;\;\; \frac{g_3^2(m_t)}{4\pi} = 0.1071 .
\ee
The couplings $g$ and $g'$ are obtained by an RG flow from their values at $\mu = M_Z$, which are given in \cite{Amsler:2008zzb}, while $g_3$ is calculated
numerically. (See \cite{Barvinsky:2009fy} and references within for details.) The pole mass matching scheme is used to set the initial conditions $\lambda_h(m_t)$ and $y_t(m_t)$. This relates the physical pole masses to the couplings in the $\overline{\mbox{{\sc ms}}}$ renormalisation scheme through the following expressions:
\bea
\lh(m_t) &=& \frac{m_h^2}{2 v^2} (1 + 2\Delta_h)\nonumber \\
y_t(m_t) &=& \frac{\sqrt{2}}{v} m_t (1 + \Delta_t)
\eea
where $\Delta_h$ and $\Delta_t$ account for radiative corrections and are given in the appendix of \cite{Espinosa:2007qp}.

For both methods we fix $\tdN = 58$ for the WMAP pivot scale (we will discuss the value of $\tdN$ in the next section), and use this to determine $\phi_{\tdN}$ via
\be{EQ:defN}
\tdN =
\int_{t}^{t_{end}} H dt
= \int^{\phi_{\tdN}}_{\phi_{end}}\frac{1}{M_p^2}\frac{\tdu}{\frac{d\tdu}{d \phi}} \left( \frac{d\chi_\phi}{d \phi} \right)^2 d\phi~,
\ee
where at the end of inflation $\phi_{end} \simeq  \sqrt{ 4 M_{p}^{2}/3 \xi_{\phi}}$.

\subsection*{RG equations and slow-roll parameters}

       The RG equations for $y_t$, $g$, $g'$, $\lh$, $\ls$, $\lhs$, $\xi_s$, $\xi_h$ and the commutator suppression factors $c_h$ and $c_s$ are presented in Appendix A. The suppression factors are inserted for each inflaton propagator in a loop. For $\phi \ll \frac{M_{p}}{\xi_{\phi}}$, $c_{\phi} \rightarrow 1$ while for $\phi \gg \frac{M_{p}}{\xi_{\phi}}$, $c_{\phi} \rightarrow 0$. Using the RG improved potential we calculate the slow roll parameters $\tdeps$ and $\tdeta$ analytically as a function of the running couplings, as discussed in Appendix A.  The spectral index is given by $n = 1-6\tdeps + 2\tdeta$. We find that the spectral index is dominated by $\tdeta$, which has only small radiative corrections. However, the tree level value of $\tdeta$ depends on $\phi_{\tdN}$, which is determined by $\tdeps$. The radiative corrections to $\tdeps$ can be large and therefore radiative corrections can have a large effect on the spectral index. The parameter $\tdeps$ is given by
\bea \label{w5x}
\tdeps & = & \frac{M_p^2}{2} \left(\frac{d\phi}{d\chi_\phi}\right)^2 \left( \frac{dU}{d\phi}\frac{1}{U} \right)^2  \\
\label{w5y}
& \approx & \frac{M_p^2}{2} \left( \frac{d\phi}{d\chi_{\phi}} \right)^2 \left(\frac{4}{\phi \Omega^2} + \frac{L_{\phi}}{\phi}\right)^2
\eea
In the second expression $\frac{4}{\phi \Omega^2}$ is the tree-level contribution and $L_{\phi}$ gives the approximate contribution from radiative corrections.  During inflation (i.e. $c_\phi = 0$), the expressions for $L_\phi$ are defined as
\bea
\label{EQ:L_hs}
16\pi^2 L_{h} \equiv   \frac{1}{\lh}\left(\frac{3g^4}{4} + \frac{3(g^2+g'^2)^2}{8} - 6y_t^4 + \frac{\lhs^2}{2} \right) - 6\lh - 2\lhs \frac{\xi_s}{\xi_h}
\eea
and
\be{EQ:L_s}
16\pi^2 L_{s} \equiv \frac{2\lhs^2}{\ls} - 2\frac{\xi_h}{\xi_s}\left(12\lh + 6y_t^2 - \frac{3}{2}\left(g^2 + g'^2\right) \right)
\ee
where the terms $\propto \xi_h/\xi_s$ and $\xi_s/\xi_h$ are likely to be subdominant. We will use the approximation \eq{w5y} to explain the main features of the results. However, the numerical results are obtained using the {\em full} expressions for $n$, $\tdeps$ and $\tdeta$, given in Appendix A.

\section{Reheating}
\label{SEC:reheating}

   In most inflation models, the uncertainty in the reheating temperature introduces a large uncertainty in the value of $N$ corresponding to the scales observed in CMB experiments
 and so in the theoretical value of the spectral index. A remarkable feature of the models we are considering is that the reheating temperature
$T_{R}$,  defined to be the effective temperature at which the inflaton energy density is equal to the energy density in relativistic particles\footnote{The temperature at which thermal equilibrium occurs may be lower than this \cite{Davidson:2000er}.}, can be determined to a high degree of precision, allowing a precise prediction of the spectral index.

     The mechanism for reheating in these models will give us a narrow range of values for  $T_R$.  This will enable us to calculate $\tdN$ to high precision -- a necessary ingredient in calculating the spectral index $n$ (\eq{EQ:SRtree}). Unlike most inflation models, the couplings of the particles to the Standard Model in Higgs Inflation and $S$-inflation are either known or well-constrained, making this calculation in principle possible to arbitrarily high precision.

However, in practice, the calculation is difficult, as non-perturbative effects are important.    Reheating in Higgs Inflation has been studied by two groups \cite{GarciaBellido:2008ab,Bezrukov:2008ut}. There has been no similar study of reheating in $S$-inflation (although some discussion was presented in \cite{Okada:2010jd}). We will show that the process is very similar to reheating in Higgs Inflation, with stochastic resonance of Higgs bosons to gauge bosons being replaced by stochastic resonance of gauge singlet scalars to Higgs bosons via the Higgs portal coupling.

  A concern in the case of $S$-inflation is that the dark matter scalars $s$ are stable and so reheating via inflaton decay is not possible. We must therefore demonstrate that reheating can occur and that there is no dangerous density of stable $s$ scalars remaining from inflation. We will show that this is the case; reheating occurs via stochastic resonance rather than decay, and any remaining energy density in $s$ oscillations will be thermalised by the dominant radiation background from reheating.

We first summarise our results. Following the approximate analytical method of \cite{Bezrukov:2008ut}, we find that $T_R$ is high for $S$-inflation:
\be{EQ:TrehS}
3\times 10^{13}\GeV \lesssim T_R \lesssim 8\times 10^{14}\GeV
\ee
and similar to $T_R$ for pure Higgs inflation \cite{Bezrukov:2008ut}
\be{EQ:TrehH}
3\times 10^{13}\GeV \lesssim T_R \lesssim 1.5\times 10^{14}\GeV.
\ee
The small range of $T_R$ for both models enables us to reasonably estimate $\tdN$. For $S$-inflation this is $58\lesssim \tdN \lesssim 61$. In this we have conservatively included an additional error $\Delta \tdN = \pm 1$ to account for possible additional uncertainties in the theoretical estimate of $T_R$, giving a classical spectral index $0.965\lesssim n_{cl} \lesssim 0.967$. A measurement of $\lhs$ through dark matter detection experiments would further increase the predictiveness of this model. For pure Higgs inflation, $57\lesssim \tdN \lesssim 60$ (also including an additional theoretical error $\Delta \tdN = \pm 1$), which gives  $0.964\lesssim n_{cl} \lesssim 0.966$. For the case of inflation in the Higgs direction including a singlet scalar, the result for $T_{R}$ will be similar to pure Higgs inflation, although a little higher (as the inflaton has an extra channel to annihilate or decay into). We next review the mechanism for reheating in Higgs Inflation before discussing the process in $S$-inflation reheating.

\subsection*{Reheating in Higgs inflation}

An analytical calculation of $T_R$ was carried out in \cite{Bezrukov:2008ut}, while in \cite{GarciaBellido:2008ab} a numerical calculation was performed.  Backreaction was discussed in \cite{GarciaBellido:2008ab}, but was not considered in \cite{Bezrukov:2008ut} since the annihilation of the gauge bosons produced via the resonance was found to be dominant in \cite{Bezrukov:2008ut}. We now further describe reheating in Higgs inflation according to \cite{Bezrukov:2008ut}, as we will follow this method to calculate $T_R$ for $S$-inflation.

      Reheating in Higgs inflation occurs through a stochastic resonance process \cite{Bezrukov:2008ut}. After inflation, the Higgs-inflaton $\chi_h$ oscillates in a quadratic ($\chi_{h}^2$) potential. The gauge boson masses are proportional to the modulus of the oscillating inflaton field. When the oscillation modulus is small during the oscillation cycle, non-relativistic gauge bosons are produced non-adiabatically; when the modulus is large, these gauge bosons rapidly decay to relativistic Standard Model particles
      \cite{Bezrukov:2008ut}. This prevents the build up of gauge bosons and backreaction. However, the expansion of the Universe causes the maximum amplitude of the Higgs oscillation to decrease, which decreases the gauge boson mass. As the decay rate of the gauge bosons is proportional to their mass, eventually the gauge bosons can no longer decay appreciably and their occupation number builds up, enabling the stochastic resonance. At this point, the energy of the inflaton is quickly transferred to the gauge bosons, which in turn quickly annihilate to relativistic fermions. Reheating via the resonant production of Higgs excitations is also possible, but the stochastic resonance to gauge bosons is expected to dominate \cite{Bezrukov:2008ut}. In Higgs Inflation, the resulting  temperature of radiation domination is $3 \times 10^{13}\GeV < T_R < 1.5 \times 10^{14}\GeV$ (where the lower limit is from reheating via the production of excitations of the Higgs) \cite{Bezrukov:2008ut}.

 The WMAP pivot scale is $k_0 = \frac{2\pi}{\lambda} = 0.002\;{\rm Mpc}^{-1} $. The number of e-foldings at which this exits the horizon is given by
\be{m2a}
\tdN = \ln{\left( \left(\frac{\tdrho_R}{\tdrho_{end}}\right)^{1/3} \left(\frac{g_0T_0^3}{g_*T_R^3}\right)^{1/3} \tilde{H}_{\tdN} \lambda  \right)},
\ee
where $g(T_R) \simeq 106.75$ for the SM, $g(T_0) \simeq 2$ and $T_0$ is the present photon temperature. This assumes that the universe is matter dominated by inflaton oscillations from the end of inflation until the moment of reheating. For the range of $T_{R}$ from Higgs inflation, and including an additional theoretical uncertainty $\tdN = \pm 1$, we obtain $57 \lesssim \tdN \lesssim 60$ for the number if e-foldings at which the pivot scale exits the horizon.

   From the expression for the spectral index \eq{EQ:SRtree}, an uncertainty
$\Delta \tdN$ corresponds to an uncertainty $\Delta n = \pm \Delta \tdN/\tdN^2$. With $\Delta \tdN = 4$ and $\tdN = 60$ this gives
$\Delta n = \pm 0.001$. Therefore, even with a conservative estimate of the uncertainty in $\tdN$, the spectral index in Higgs inflation can be calculated to an accuracy  $\pm 0.001$. This can be improved simply by improving the accuracy of the reheating temperature calculation; in principle, there are no obstacles to an arbitrarily precise calculation of the spectral index in these models.

\subsection*{\texorpdfstring{Reheating in $S$-inflation}{Reheating in S-inflation}}
\label{SEC:give}
Reheating in $S$-inflation occurs by the production of Higgs bosons
via the coupling of $s$ to $H$, which can then decay to the particles of the Standard Model. Since $s$ is stable, any leftover inflaton density can only be transferred to thermal radiation through scattering with Higgs bosons and $s$ bosons in the thermal background. We therefore need to check that the inflaton is a subdominant component of the Universe when its oscillation is small enough that it effectively oscillates in a quartic potential ($\propto \chi_{s}^4 \approx s^4$). In this case, any residual oscillating inflaton density will eventually be thermalised by scattering from the dominant thermal background. Taking a conservative approach, we therefore require that reheating should occur during the early quadratic ($\chi_{s}^2$) potential stage, where $\chi_{s}$ is the canonically normalised Higgs field. (This is conservative as it still may be possible to have reheating once $\chi_{s}$ is oscillating in the quartic potential.) Reheating  occurs via stochastic resonance to Higgs bosons through $\frac{\lhs}{2}s^2|H|^2$. The process is very similar to the case of reheating in Higgs Inflation, with the gauge boson final state particles replaced by Higgs bosons and the gauge couplings replaced by $\lhs$.  We therefore apply same method as used in the case of Higgs Inflation \cite{Bezrukov:2008ut} to estimate the reheating temperature. Our calculation is outlined in Appendix B.

The key points are as follows. Once the $\chi_s$ oscillation amplitude becomes small enough to allow stochastic resonance to occur, the energy in the $\chi_{s}$ oscillations rapidly transfers to the Higgs bosons, which in turn decay to SM fermions (primarily top quarks, since the $\chi_s$ oscillation amplitude gives the Higgs bosons a large mass). This causes the $\chi_s$ amplitude to rapidly decrease, such that $\Omega \rightarrow 1$ and the $s$ field becomes canonically normalised in an $s^4$ potential. Provided that the radiation energy density from Higgs boson decay is dominant at this time, the $s$ oscillations and radiation will subsequently decrease as $a^{-4}$ and the radiation energy density will therefore remain dominant at least until the time at which the $s$ mass term comes to dominate the $s$ oscillations. Provided that the thermal radiation background can thermalise the subdominant $s$ oscillations before this time, there will be no stable $s$ density remaining
from inflation. In this case $s$ dark matter will be a purely thermal relic density.

In fact, there are two possible mechanisms for reheating in $S$-inflation. The first is through stochastic resonance to Higgs bosons, which gives
\be{Tmax}
T_R \approx 3\times 10^{13} \left(\frac{\ls}{\lhs^2}\right)^{1/4}\GeV  ~.
\ee
This requires $\ls>0.25 \lhs$ in order for reheating to complete before the potential changes from quadratic to quartic. Imposing a bound{\footnote{There are constraints due to vacuum stability and perturbativity of the couplings (see \cite{Lerner:2009xg}) and from the requirement that $\lhs$ must be large enough to achieve the correct relic density of dark matter.}} $\frac{\ls}{\lhs^2}\lesssim 10^6$ gives
\be{ee2}
3 \times 10^{13} \GeV < T_R < 8 \times 10^{14} \GeV.
\ee
The lower bound is similar to that for Higgs inflation \cite{Bezrukov:2008ut}. The upper bound for $S$-inflation is higher because of the freedom in $\ls$ and $\lhs$.

The second mechanism is reheating via the production of excitations of the inflaton itself, which subsequently annihilate to Higgs bosons. This gives
\be{T_R_ex}
T_R \approx 9 \times 10^{13} \ls^{1/4} \GeV
\ee
and requires $\ls>0.019$ to complete before the potential becomes quartic.

\begin{figure}[tb]
                    \centering
                    \includegraphics[width=0.4\textwidth, angle=0]{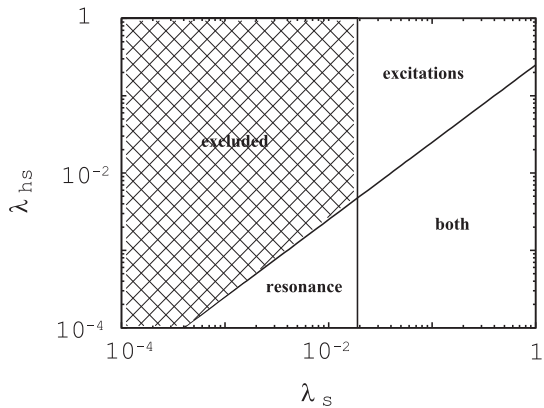}
                    \caption{\footnotesize{Showing the regions of the $\ls$ and $\lhs$ parameter space allowed by the constraints of reheating. The shaded region is excluded under the conservative assumptions described in the text. In the majority of the region marked `both', reheating via stochastic resonance is expected to occur first and therefore to dominate.}
                    \label{couplim}}
\end{figure}

A region of the parameter space is excluded due to the conditions on the couplings, as shown in \fig{couplim}. However, we are rather conservative in our estimates since some reheating is still likely to occur once the oscillations become quartic. With $T_R$ given by \eq{ee2} we find $59 \lesssim \tdN \lesssim 60$. Including a conservative estimated theoretical error of $\pm 1$ on $\tdN$ we arrive at
\be{EQ:Nresult}
58 \lesssim \tdN \lesssim 61.
\ee
Thus we expect the uncertainty in $\tdN$ and the resulting error in the theoretical estimate of the spectral index to be similar to the case of Higgs Inflation, $\Delta n = \pm 0.001$, which could be improved with more careful calculations.

\subsection*{Thermalisation of the remaining inflaton oscillations}

Because $s$ cannot decay, it is necessary to check whether there is any energy density remaining in the inflaton oscillations after reheating.
From the end of inflation (when resonant reheating occurs) until the time when the mass term in the $s$ potential comes to dominate the $s^4$ term, the radiation background will dominate the inflaton oscillations. The SM radiation background will thermalise well before the $s^2$ term comes to dominate. To check that the remaining energy in the $s$ oscillations is thermalised, we can simply check that the energy in the $s^2$ oscillations, equivalent to a density of zero momentum $s$ particles, is thermalised before the $s^2$ oscillations dominate the thermal background.

The condition for sufficiently rapid thermalisation
of the inflaton oscillations in the $s^2$ potential is $\Gamma > H$, where $\Gamma$ is the scattering rate of zero-momentum $s$ particles from particles in the thermal background (which also includes thermal $s$ particles) and $H$ is the Hubble parameter. We will only consider scattering with the Higgs bosons, as this will prove sufficient to demonstrate thermalisation.
(Zero-momentum $s$ particles could also annihilate with thermal background $s$ with a similar rate.) The scattered $s$ particles will achieve a thermal energy by subsequent scatterings and will achieve a thermal equilibrium number density via annihilation to SM particles, provided scattering occurs at a temperature greater than the freeze-out temperature of the $s$ scalars.

  The scattering rate of zero-momentum $s$ scalars by Higgs bosons is $\Gamma \approx n \sigma v$, where $n$ is the number density of thermal background Higgs bosons,
  $n = \frac{1.2}{\pi^2} g_H T^3$  with $g_H =4$, $v = 1$ is the velocity of the relativistic Higgs particles and $\sigma$ is the scattering cross-section,
\be{EQ:cross}
\sigma \simeq \frac{\lhs^2}{96\pi \mso T}.
\ee
Assuming $\rho_{total} \simeq \rho_{rad}$, $H \simeq 3.4 T^2/M_p$. Therefore thermalisation of the background will occur if
\be{hhuhh}
\frac{\lhs^2}{24\pi^2} \frac{T_m^2}{\mso} \gtrsim \frac{3.4 T^2}{M_p}.
\ee
This is satisfied if
\be{mson}
\mso \lesssim 1.0 \times 10^{-4} \lhs^2 M_p~.
\ee
This is easily satisfied, as we expect $\mso$ to be less than a few TeV. This means that any inflaton energy density remaining in the $s$ oscillations will be scattered and thermalised as soon as the oscillation amplitude enters the $s^2$ oscillation regime, if not before. A key point is that since the energy density of the Universe is dominated by radiation at the onset of $s^4$ oscillations, there is sufficient energy to completely thermalise the $s^2$ oscillations. Of course, it is likely that thermalisation of the oscillations will occur even earlier; our calculation is simply a check that thermalisation of the $s$ oscillations can safely occur. Thus there is no danger of
a residual dark matter density in $s^2$ oscillations; $s$ dark matter is entirely from thermal freeze-out of $s$ particles.

\section{\texorpdfstring{Distinguishing $S$-inflation from Higgs-inflation through observations}{Distinguishing S-inflation from Higgs-inflation through observations}}
\label{SEC:compresults}

    We next present the main result of our paper, which is the prediction of the spectral index $n$ and tensor-to-scalar ratio $r$ as a function of $m_h$ for the three variants of the model.
Compared to pure Higgs inflation, the gauge singlet variants of the model introduce two new parameters $\ls$ and $\lhs$. The second can potentially be fixed by experiment. There will also be a minimum value of $\lhs$ below which the $s$ particles cannot account for the observed density of dark matter. We examine all values of $\ls$ and $\lhs$ at $\mu = m_{t}$ where the magnitude of all couplings (except $\xi_{\phi}$) remain perturbative (we generously take this to be $<100$) up to the scale of inflation. We have not imposed any limits on the perturbativity of the potential in the non-inflationary direction. Along with stronger limits in the inflationary direction, this would impose stronger bounds on the permitted combinations of $m_h,~\ls$ and $\lhs$. Our aim here is only to compare the behaviour of $n$ and $r$ as a function of $m_h$.

We find that the spectral index is well approximated (in most of the parameter space) by considering only the tree-level contribution to $\tdeta$
\be{EQ:clasn}
n \simeq 1 + 2\tdeta \simeq  1- \frac{8}{3}\frac{M_p}{\xi_{\phi} \phi_{\tdN}^2}.
\ee
Thus, the shape of $n$ is determined by $\xi_{\phi} \phi_{\tdN}^2$, where this is determined by the radiatively-corrected $\tdeps$  through the normalisation to the COBE data (\eq{EQ:newWMAP}) and through the integration to get $\tdN$, \eq{EQ:defN}. If radiative corrections cause $\tdeps$ to increase above its classical value, we would expect $\xi\phi_{\tdN}
^2$ to be larger for a fixed $\tdN$. This means that the magnitude of $\tdeta$ is increased (i.e. becomes less negative) and so $n$ is increased.

We will use the quantity $L_{\phi}$ as introduced in \sect{radcor}, which is approximately the part of $\frac{\phi}{U}\frac{dU}{d\phi}$ due to radiative corrections, to explain how $n$ deviates from its classical value. Large $L_{\phi}$ corresponds to large radiative corrections to $\tdeps$, which corresponds to larger $n$, as described above. $L_{h}$ and $L_s$ are given by \eq{EQ:L_hs} and \eq{EQ:L_s}.

\subsection*{\texorpdfstring{Spectral Index as a function of $m_{h}$}{Spectral Index as a function of Higgs mass}}
\begin{figure}[tb]
  \centering
    \subfloat[\footnotesize{Spectral index $n$. The points curving downwards correspond to the $H$-direction.}]
    {\label{FIG:n5b} \includegraphics[width=0.32\textwidth, angle=270]{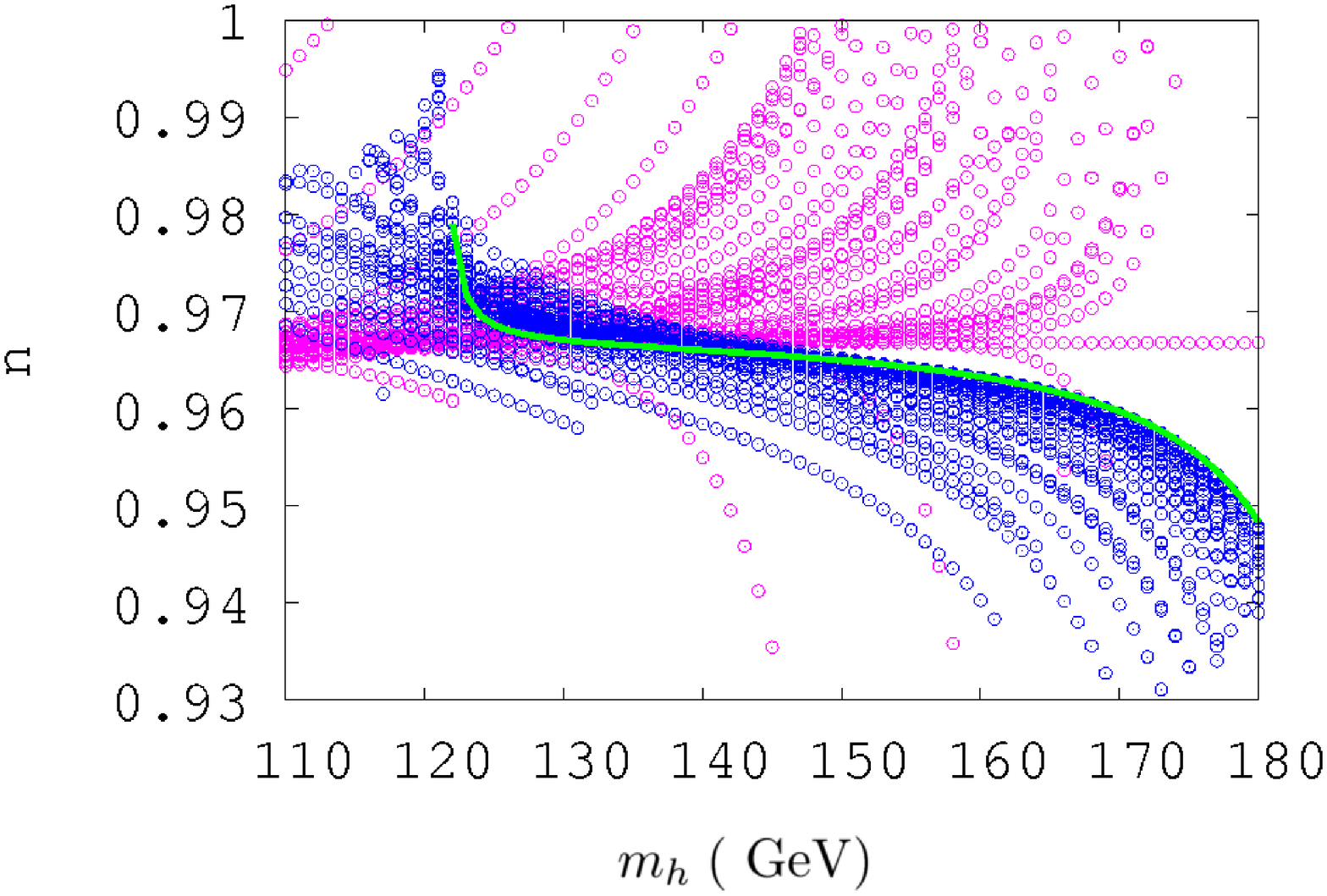}}
    \hspace{0.3in}
    \subfloat[\footnotesize{Tensor to scalar ratio $r$. The points curving downwards correspond to the $H$-direction.}]
    {\label{FIG:r5} \includegraphics[width=0.32\textwidth, angle=270]{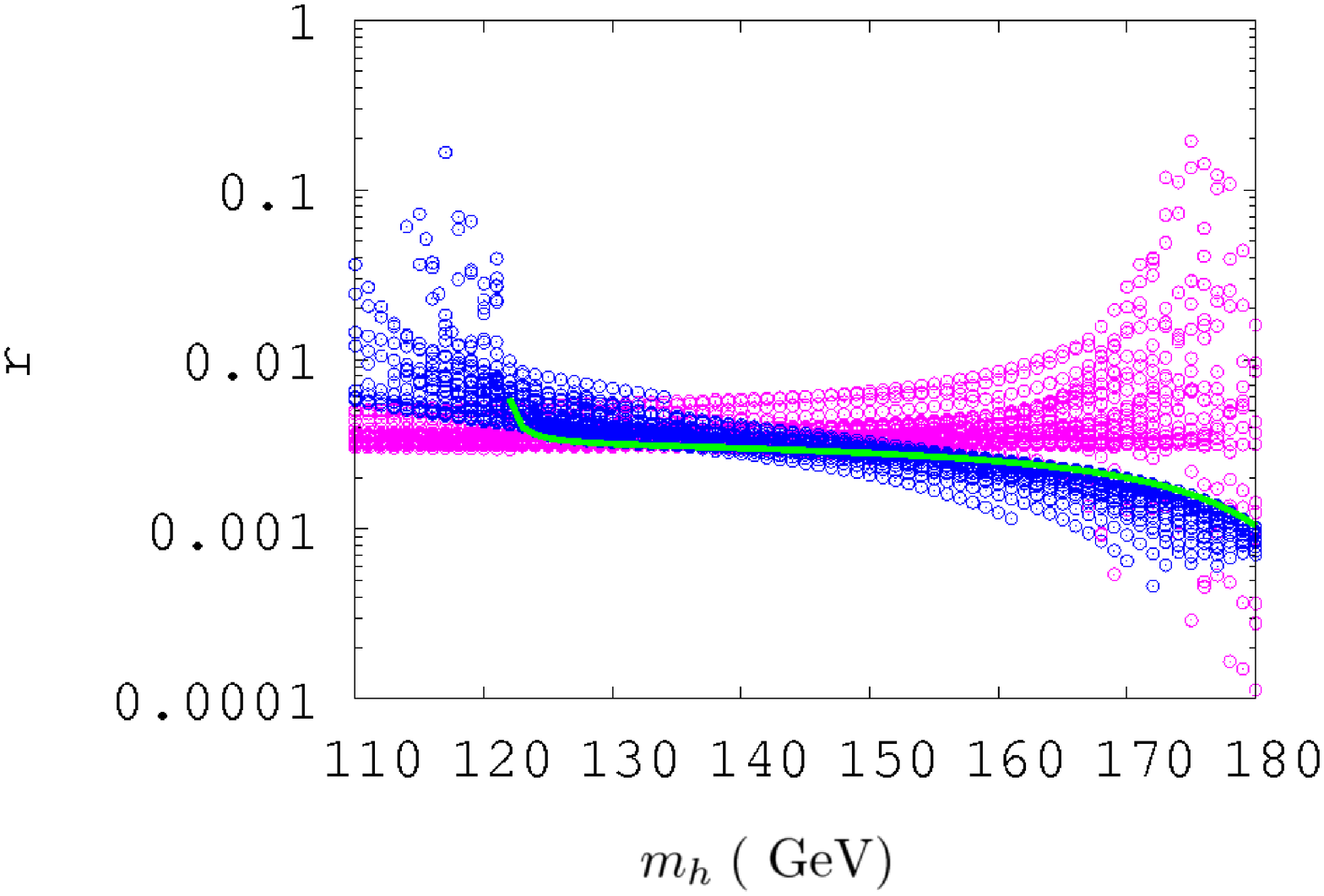}}
  \caption{\footnotesize{Spectral index $n$ and tensor to scalar ratio $r$ versus Higgs mass $m_h$ for  inflation in the S-direction (pink circles), for inflation in the H-direction (blue circles) and for pure Higgs inflation (solid green line). Couplings have been varied by increments of 0.1 and required to remain less than $100$.}}
\end{figure}

  We first present our main result, which is the very different behaviour of the spectral index and tensor-to-scalar ratio as a function of Higgs mass in $S$-inflation as compared with Higgs Inflation models. We then explain the behaviour of the spectral index as a function of the Higgs mass and model couplings in the different models.

We show in \fig{FIG:n5b} the approximate range of $n$ as a function of $m_{h}$ for each model. In these plots $\ls$ and $\lhs$ are allowed to take any values that are multiples of 0.1.    Introducing perturbativity and vacuum stability bounds would cause these areas to somewhat decrease in size. We see that there is a striking difference between the two models at larger values of $m_h$. At $m_h \gtrsim 150 \GeV$, the possible values of $n$ are very different{\footnote{The small number of $s$-direction points with $n<n_{cl}$ are likely to be ruled out by perturbativity and stability constraints on the couplings.}}. There is more overlap at lower $m_h$ --- for $125 \GeV \lesssim m_h \lesssim 135 \GeV$, it appears unlikely that $n$ could discriminate between the models. Of course, a measurement of $\lhs$ will leave only one free parameter ($\ls$), reducing the number of points available.

We show in \fig{FIG:r5} the equivalent figure for $r$, with the same restrictions on $\ls$ and $\lhs$ as above. We see that $r$ is in general low $(r\lesssim 0.02)$ --- although for large $m_h$ ($s$-direction) and small $m_h$ ($H$-direction), it can take values which are only just within the current WMAP limit $r < 0.22$. Thus there is a small chance that $r$ may be detectable by Planck, although these extreme points may be excluded when full stability and perturbativity constraints are applied to the models{\footnote{These constraints should be applied in {\em both} directions ($h$ and $s$), regardless of the direction of inflation, and should be applied at least to a scale just above the scale of inflation.}}.

   As we have shown, the error on the theoretical value of $n$,  due to uncertainty in the reheating temperature,  is $ \Delta n = \pm 0.001$ or less.
We expect the Planck experiment to measure $n$ to a 2-$\sigma$ accuracy of $\pm 0.0005$. Therefore if Planck should find $n$ significantly larger than 0.967 while LHC finds a Higgs with mass larger than $135 \GeV$, $S$-inflation will be compatible with the observations but Higgs inflation will be ruled out. If the spectral index $n$ is measured to be significantly less than 0.965 then Higgs inflation will be compatible with observations while $S$-inflation would be ruled out.

     Thus a combination of LHC data and Planck data can determine which weak scale inflation scenario is viable. The possible measurement of $\lhs$ in direct detection dark matter experiments (and, if $m_{s} < m_{h}/2$, at the LHC), combined with a Planck measurement of $n$ larger than (but close to) 0.967 and negligible $r$, would be strong evidence in support of $S$-inflation.

   We next explain in more detail the behaviour of the spectral index as a function of $m_{h}$ in the cases where the Higgs is the inflaton and $s$ is the inflaton.

\subsection*{Spectral index with the Higgs as inflaton}
\begin{figure}[tb]
  \centering
    \subfloat[\footnotesize{Inflation in the Higgs direction with $\ls = 0$ and varying $\lhs$: $\lhs = 0$ (pure Higgs inflation; red and furthest right), $\lhs = 0.1$ (green), $\lhs = 0.3$ (pink) and $\lhs = 0.5$ (blue, furthest left).}]
    {\label{FIG:n1} \includegraphics[width=0.3\textwidth, angle=270]{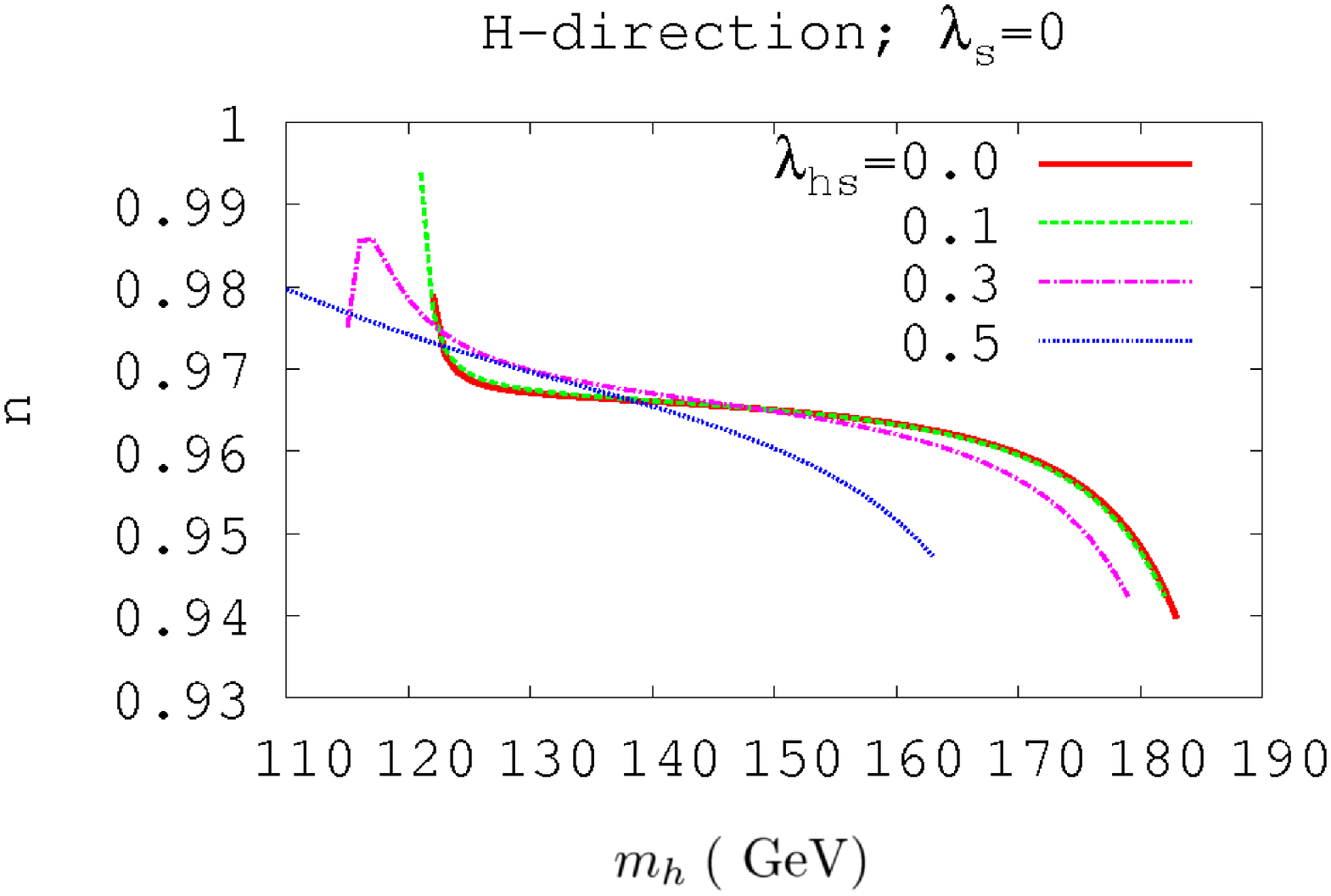}}
    \hspace{0.3in}
    \subfloat[\footnotesize{Inflation in the Higgs direction, with $\ls = 0.0$ (solid red) and $\ls = 0.15$ (green dashed).}]
    {\label{FIG:n2} \includegraphics[width=0.3\textwidth, angle=270]{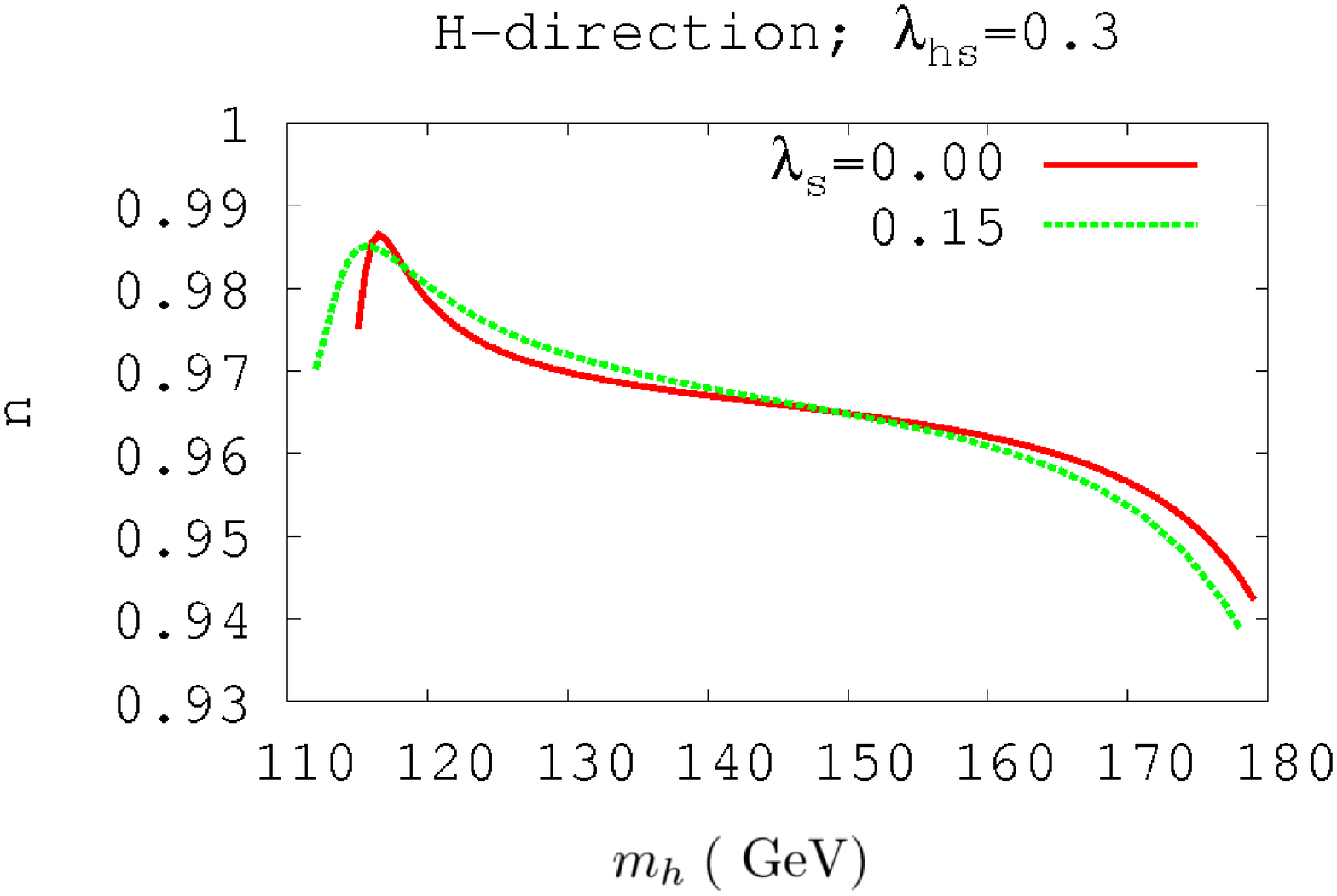}}\\
    \subfloat[\footnotesize{Inflation in the S-direction with $\lhs = 0.0$: $\ls = 0.01$ (solid red) and $\ls = 0.25$ (green dashed). `Pure' Higgs inflation is shown for comparison (solid black).}]
    {\label{FIG:n3} \includegraphics[width=0.3\textwidth, angle=270]{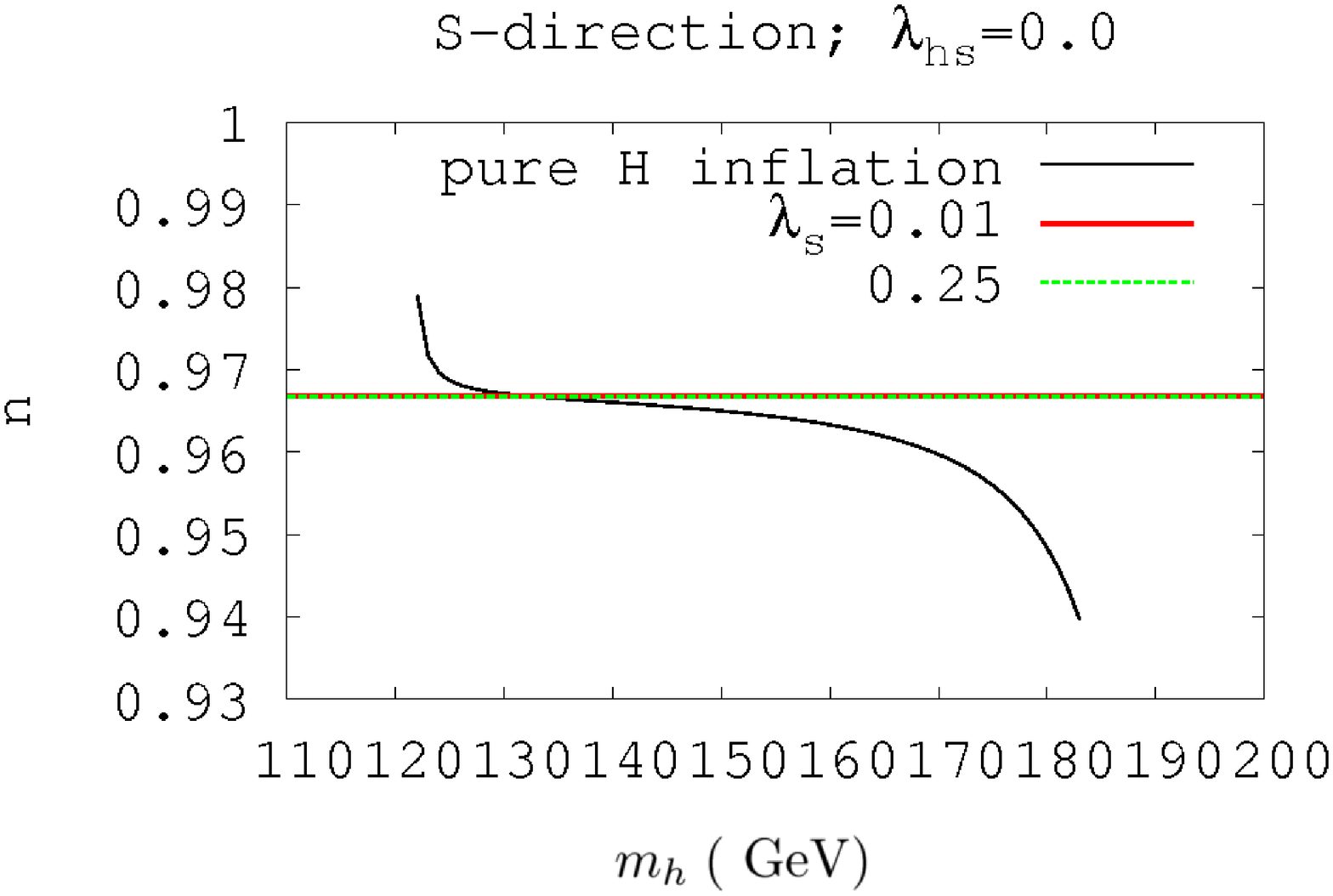}}
    \hspace{0.3in}
    \subfloat[\footnotesize{Inflation in the $S$-direction, with $\ls = 0.1$ (solid) and $\ls = 0.01$ (dashed). Various values of $\lhs$ are shown: $\lhs = 0.0$ (black, horizontal) and from right to left, $\lhs = 0.01$ (blue), $\lhs = 0.1$ (red), $\lhs = 0.3$ (green) and $\lhs = 0.5$ (pink).}]
    {\label{FIG:n4} \includegraphics[width=0.3\textwidth, angle=270]{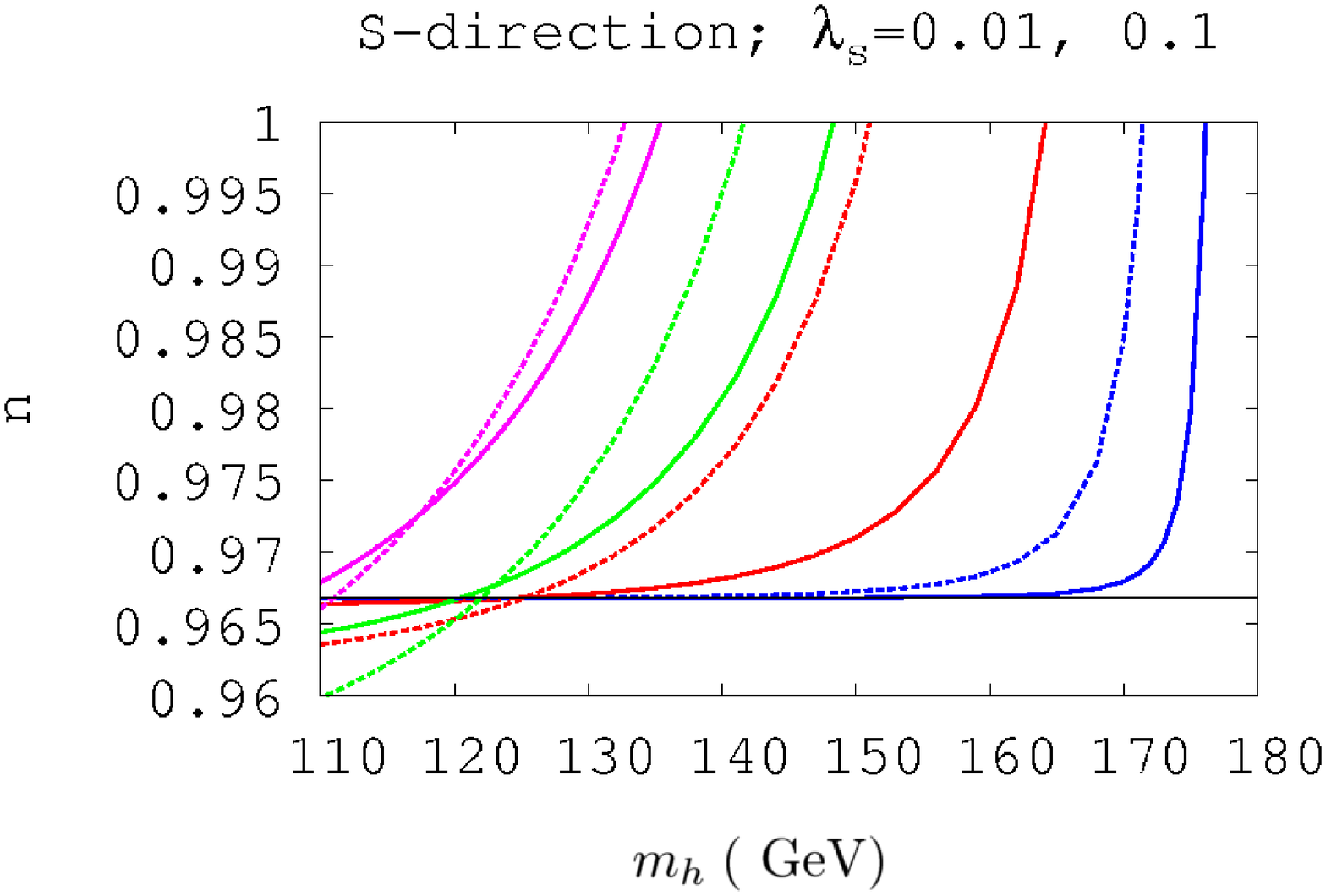}}\\
  \caption{\footnotesize{Spectral index $n$ versus Higgs mass $m_h$ for different versions of the model}  \label{mainresults} }
\end{figure}

Firstly, we consider the effect of $\lhs$ on Higgs inflation and so set $\ls(m_t) = 0$ (but allow for its running). The results are shown in \fig{FIG:n1} for $\lhs = 0$ (red; pure Higgs inflation), $\lhs = 0.1$ (green), 0.3 (pink) and 0.5 (blue). The shape of the curves and range of $n$ for all $\lhs$ are quite similar (for $m_h > 122\GeV$), but two main features can be seen: (i) as $\lhs$ increases, the curves shift to the left, also shifting the range of $m_h$ and (ii) there is a turnover at low $m_h$ for $\lhs = 0.3$. These features are explained below.

\noindent (i) A larger $\lhs$ increases $\beta_{\lh}$ (\eq{EQ:beta_h}), giving a larger $\lh$ for a given $m_h$. At large $m_h$, $L_H \propto -6\lh$ (\eq{EQ:L_hs}), and so $n$ falls faster with larger $\lhs$. This explains the shift to the left as $\lhs$ increases. The range of $m_h$ accessible to this model is shifted downwards as $\lhs$ increases. A larger $\lhs$ can prevent $\lh$ becoming negative at low values of $m_h$. Large $\lhs$ causes the couplings to grow faster and the perturbativity limit to be reached at a lower $m_h$. Thus, the range of $m_h$ is shifted.

\noindent (ii) The turnover at low $m_h$ (seen for the case $\lhs = 0.3$ in \fig{FIG:n1}) is due to the term in $L_H \propto \frac{\xi_s}{\xi_h}$. Small $\lh$ means that $\xi_h$ is small (from normalisation, \eq{COBEnorm}). Large $\lhs$ gives a large running of $\xi_s$ (remember that $\xi_s(m_t) = 0$) so the ratio $\frac{\xi_s}{\xi_h}$ is of ${\cal O}(1)$ at the scale of inflation{\footnote{However, with $\xi_s$ and $\xi_h$ of similar magnitudes, we can no longer assume that inflation is in the $H$-direction. Therefore this is perhaps not a valid region of the parameter space --- however we do not impose any constraint on $\frac{\xi_s}{\xi_h}$ in this chapter.}}. Thus there is a large negative contribution (last term of \eq{EQ:L_hs}) almost balancing the positive term $\propto \frac{1}{\lh}$.

As shown in \fig{FIG:n2}, $\ls$ has a steepening effect on $n$. However, this is a relatively small effect for $\lhs(m_t) = 0.3$. Above about $\ls = 0.15$, we reach the perturbativity limit of $\ls$. The steepening is due to $\ls$ causing $\lhs$ to increase, therefore exaggerating the effects of \fig{FIG:n1} further. The last term in \eq{EQ:L_hs}, $-2\lhs \frac{\xi_s}{\xi_h}$, may also play a role. Increasing $\ls$ increases $\xi_s$, which may give some contribution to the decrease in $n$ at larger $m_h$. For $\lhs(m_t) = 0.1$ (not shown) we find that increasing $\ls$ has a negligible effect on $n$ ($\ls$ becomes non-perturbative at $\ls(m_t) \sim 0.3 - 0.4$ in this case).

We conclude that introducing a real singlet scalar to the model of Higgs inflation can affect the spectral index prediction, increasing it at low $m_h$ and decreasing it at higher $m_h$. This effect is controlled mainly by the magnitude of $\lhs$ which is, in principle, measurable. The addition of $\lhs$ also changes the range of $m_h$, decreasing both upper and lower limits. Negative values of $\lhs$ are allowed and give similar results. The spectral index is insensitive to $\lambda_s$.

\subsection*{\texorpdfstring{Spectral index in $S$-inflation}{Spectral index in S-inflation}}

    With $\lhs(m_t) = 0$ we see that the spectral index does not vary noticeably with $m_h$ --- see \fig{FIG:n3} (where we show the pure Higgs inflation case for comparison). This is as we would expect, since with $\lhs = 0$, the model is completely decoupled from the Higgs sector. We also see that $n$ does not vary with $\ls$ (if $\ls$ is increased much further than shown in the figure, it reaches its perturbativity limit). This is because the deviation of $n$ from its classical value is determined by $L_S$ (\eq{EQ:L_s}) which is proportional to $\lhs^2 = 0$.

    With non-zero $\lhs$, we show the results for $\ls(m_t) = 0.1$ (solid) and  $\ls(m_t) = 0.01$ (dashed) in \fig{FIG:n4}. We see that increasing $\lhs$ has a dramatic effect on $n$. This is because $L_S \propto \frac{\lhs^2}{\ls}$. Large $\left|\lhs\right|$ (at inflation scale) will therefore cause $L_S$ to increase, giving larger $n$. Smaller $\ls$ gives larger $L_S$, increasing $n$ further, although the effect of $\lhs$ is dominant.

      There is a significant sensitivity to the value of $\ls$. In $S$-inflation it is possible, in principle, to measure all the unknown couplings with the exception of $\ls$. Thus, although the qualitative behaviour of $n$
as a function of $m_{h}$ can be known, there is a fundamental obstacle to making a precise prediction of $n$ for a given $m_h$ in $S$-inflation. However,  with a reasonable estimate of the value of $\ls$, such as $\ls \approx \lambda_{h}$, we can estimate the expected value of $n$. In addition, constraints on $\ls$ from vacuum stability and perturbativity of the potential will also constrain the
allowed range of $n$.

\section{Conclusions}
\label{SEC:conclusions}

     The Standard Model and its minimal extension to include a real gauge singlet scalar dark matter particle can account for inflation if the scalar fields are non-minimally coupled to gravity with a large dimensionless coupling $\xi_{\phi}$.
In this paper we have considered the precision predictions of these models for $n$ and $r$ as a function of Higgs mass.
In order to ensure a consistent calculation and so avoid the contradictions found in the existing results for the spectral index in Higgs Inflation, we have recomputed the RG improved potential for Higgs Inflation and $S$-inflation using the same method. The addition of the gauge singlet scalar allows inflation along both the Higgs direction (Higgs Inflation) and the singlet
direction ($S$-inflation). We have shown that these alternatives can be clearly distinguished by the deviation of the spectral index from its classical value as a function of $m_h$ if the Higgs mass is sufficiently large, $m_h \gtrsim 135 \GeV$. In $S$-inflation,
quantum corrections to the inflaton potential cause $n$ to increase with $m_{h}$, whereas in Higgs Inflation (both in the SM and in its singlet scalar extension), $n$ decreases with increasing $m_h$. The theoretical prediction of $n$ is accurate
to $\pm 0.001$, so observation of $n$ by Planck, combined with an LHC determination of $m_h$, has the potential to distinguish which of the models is consistent with data. For smaller $m_h$, both Higgs Inflation and $S$-inflation can produce values of $n$ which are larger than the classical value, so in this case it is more difficult to distinguish the models, although since pure Higgs Inflation can, in principle, make arbitrarily accurate predictions for $n$ as a function of $m_h$, it would still be possible to exclude pure Higgs Inflation. We find that the additional singlet scalar also has a small but potentially observable effect on the predictions for $n$ and $r$ in the Higgs direction. Finally, all the models we have considered predict a small value of the tensor to scalar ratio $r \lesssim 0.02$ over most of the parameter space.

      In most inflation models, the main theoretical uncertainty in the prediction of $n$ is due to the reheating temperature. A striking feature of Higgs Inflation and its singlet variants is that the reheating temperature can be determined very precisely, since the relevant
model couplings are either completely determined experimentally (in the case of pure Higgs Inflation) or determined up to the unknown values of $\ls$ and $\lhs$ (in $S$-inflation).  Although there is a somewhat greater uncertainty in the case of $S$-inflation, with plausible assumptions based on the requirement of thermal relic $s$ dark matter, a very small range of $T_R$ can be determined. We find that $n$ can be predicted up to an error $\Delta n = \pm 0.001$. Therefore if Planck can determine $n$ to an accuracy of
$\pm 0.0005$, then it should be possible to distinguish whether Higgs Inflation or $S$-inflation is consistent with observation  for a range of $m_h$.

                   In the case of Higgs inflation, a precise test of the model is possible since $n$ can be
exactly calculated as a function of $m_h$. In the case of $S$-inflation, the $s$ self-coupling $\ls$ probably cannot be directly measured but it can significantly modify the value of $n$. Therefore a precise prediction of $n$ as a function of $m_h$
may not be possible. However, the magnitude of the positive deviation from the classical value of $n$ can be estimated under plausible assumptions regarding the magnitude of $\ls$, including full consideration of the perturbativity constraints. Combined with observation of $s$ dark matter in direct detection experiments, and possible
confirmation of $s$ particles by LHC in the case where $m_s < m_h/2$, this would provide strong support for $S$-inflation.

Our results are dependent upon a number of assumptions being correct. Firstly, we assume that the effective potential is not modified by the physics that unitarizes Higgs scattering due to the non-minimal coupling to $R$. This is most likely to be true if the scale of perturbative unitarity violation, $E \sim \Lambda = M_{p}/\xi_{\phi}$, is in fact the scale at which strong coupling unitarizes the scattering process. In this case, no modification of the action is necessary. Secondly, in keeping with all previous analyses of the effective potential, we assume that the effect of the non-minimal coupling on the RG improved effective potential is fully taken into account by the suppression of the inflaton propagator in the Jordan frame at $\phi > \Lambda$. The validity of both of these assumptions merits further investigation.

    Higgs Inflation and its variants generally assume that corrections due to Planck scale-suppressed operators, such as may be expected from the UV completion of the theory, do not significantly modify the
predictions of the model. Such corrections could modify both the inflaton potential and  the non-minimal coupling. Whether these corrections can modify the model predictions for $n$ and $r$ will depend on the form of the UV completion. If the non-renormalizable potential terms in the Lagrangian produce physical interactions with strength determined by the Planck scale, then the corresponding Lagrangian operators will have additional factorial suppression factors which will ensure that their contribution to the spectral index is negligible. (Such corrections can be comparable to 1-loop corrections in the absence of factorial suppression \cite{Bezrukov:2008ut}.) Planck corrections to the non-minimal coupling will also be negligible if the full UV completion of the non-minimal coupling is an expansion of the form $\xi f(|H|^2/M_{p}^2)R$, where $f$ is expanded in $|H|^2/M_{p}^2$.  Since the effect of Planck-scale corrections on the effective potential is difficult to determine, the best strategy is to test the models experimentally.

In conclusion, it is possible that inflation and thermal relic WIMP dark matter might be understood entirely in terms of the Standard Model and its simplest extension to include a singlet scalar dark matter particle. We have shown that the nature of inflation in such models may be determined using forthcoming LHC and Planck data for a range of Higgs mass and spectral index.

\section*{Acknowledgements}

This work was supported by the Academy of Finland grant 1131454 and by the European Union through the Marie Curie Research and Training Network "UniverseNet" (MRTN-CT-2006-035863).

\section*{Appendix A: Radiative corrections and slow roll parameters}
\renewcommand{\theequation}{A-\arabic{equation}}
 \setcounter{equation}{0}

In this appendix, we detail the equations firstly of the RG improved effective action (including the suppression of the scalar commutator) and secondly of the slow roll parameters.

\subsection*{(i) The modified commutator}
In our calculation of the RG-improved effective potential we have applied the prescription that the effect of the non-minimal coupling is to suppress the inflaton propagator in the Jordan frame when calculating Feynman rules. We briefly review the argument for this method.

The non-minimal coupling to gravity in the Jordan frame means that the scalar field propagator (proportional to the commutator $[\phi(\vec{x}),\dot{\phi}(\vec{y})]$) is modified. The technique for calculating this modification was introduced by \cite{Salopek:1988qh} and first applied to the case of Higgs inflation by \cite{DeSimone:2008ei}.

The effective potential is calculated in the Jordan frame, using the RG equations derived in the Jordan frame, where the fields are canonically normalised.
The scalar field commutator
\bea
\label{comE}
[\phi(\vec{x}), \pi(\vec{y})] = i\,\hbar\delta^3(\vec{x} - \vec{y}).
\eea
is computed by first transforming the action to the Einstein frame, where the gravitational term is minimal and the canonical momentum $\pi$ can be calculated. This is given by
\bea
\label{EQ:canmomE}
\pi &=& \frac{\partial {\cal L}}{\partial(\partial_0 \phi)} = \sqrt{-\tilde{g}}\left(\tilde{g}^{0\nu} \left( \frac{d\chi}{d\phi} \right)^2  \partial_\nu \phi \right)
 =  \Omega^2 \left( \frac{d\chi}{d\phi} \right)^2\sqrt{-g}\left(g^{0\nu}\partial_\nu \phi \right) .
\eea
where $\chi$ is the rescaled field with canonical normalisation in the Einstein frame.
 Using the commutator \eq{comE} and rearranging, the result is
\bea
\label{comm2}
[\phi(\vec{x}),\dot{\phi}(\vec{y})] &=& \frac{1}{\Omega^2}\left( \frac{d\chi}{d\phi} \right)^{-2}  i\hbar \delta^{(3)}(\vec{x}- \vec{y}) \nonumber \\
& = & i\hbar\, c_\phi\, \delta^{(3)}(\vec{x}- \vec{y}).
\eea
Therefore all propagators for scalars $\phi_i$ are in principle suppressed by factors $c_{\phi_i}= \frac{1}{\Omega^2}\left( \frac{d\chi_i}{d\phi_i} \right)^{-2}$. However, the only scalar propagator which is suppressed is the non-minimally coupled inflaton field which has a large expectation value, since for all other scalar fields
$\left( \frac{d \chi}{d \phi_{i}} \right)^2 = \frac{1}{\Omega^2}$. (This can be seen from \eq{e4}.)
In this case the commutator should be understood as applying to perturbations about the background inflaton field.
The suppression factor for the inflaton is given by \cite{Salopek:1988qh,DeSimone:2007pa}
\bea \label{EQ:supressionfac}
c_{\phi} = \frac{1 + \frac{\xi_{\phi} {\phi}^2}{M_p^2}}{1 + (6\xi_{\phi} + 1)\frac{\xi_{\phi} {\phi}^2}{M_p^2}}.
\eea
It is important to emphasise that in the case of the Higgs as inflaton, only the physical Higgs scalar $h$ has a suppression factor, not the Goldstone bosons in $H$. In our calculations, although suppression factors for both $h$ and $s$ are included in the RG equations, the suppression factor for the field which is {\em not} the inflaton is set to 1.

\subsection*{(ii) RG equations}

We use the two-loop RG equations for all the SM couplings, inserting the suppression factor $c_\phi$ for each non-minimally coupled inflaton $\phi$ running in a loop, as described in the previous section. For any coupling $\lambda$ we have
\be{ww5} \frac{d\lambda}{dt} = \beta_\lambda
\ee
for inflation in the $s$ direction (since $\gamma_s = 1$), and
\be{ww4} \frac{d\lambda}{dt} = \frac{\beta_\lambda}{1+\gamma_H}
\ee
for inflation in the $h$ direction, where $t = \ln{\frac{\mu}{m_t}}$ and
\bea
\label{EQ:gamma}
\gamma_H &=& -\frac{1}{16\pi^2}\left(\frac{9g^2}{4} + \frac{3g'^2}{4} - 3y_t^2\right) - \frac{1}{(16\pi^2)^2}\left(\frac{271}{32}g^4 - \frac{9}{16}g^2g'^2 - \frac{431}{96}c_hg'^4 \right.\nonumber \\
& & \left.- \frac{5}{2}\left(\frac{9}{4}g^2 + \frac{17}{12}g'^2 + 8g_3^2\right)y_t^2 + \frac{27}{4}c_hy_t^4 - 6c_h^3\lh^2\right).
\eea

The SM one- and two-loop equations can be found in \cite{DeSimone:2008ei} and \cite{Espinosa:2007qp}. The RG equations for the scalar couplings can be obtained{\footnote{These were also obtained by \cite{Clark:2009dc}, although there is a minor difference between our work and \cite{Clark:2009dc} relating to a factor of $c_h$ in the $\xi_h$ RG equation. The term in $\beta_\xi$ proportional to $\lh$ is suppressed by a factor $(1+c_h^2)$ in Eq.\ (A2) of \cite{Clark:2009dc}, but $(1+c_h)$ in our work \cite{Lerner:2009xg}. We believe that the latter is correct because only one physical Higgs $h$ runs in the corresponding loop. In practice  the effect of the difference between these is negligible.}} using the technique detailed in \cite{Machacek}, as in \cite{Lerner:2009xg}. The one-loop $\beta$-functions for the scalar couplings are
\bea \label{EQ:beta_h} 16\pi^2\beta_{\lh}^{(1)}  =     - 6y_t^4 + \frac{3}{8}\left(2g^4 + \left(g^2+g'^2\right)^2\right) + \left(-9g^2 -3g'^2 + 12y_t^2\right)\lh +\left(18c_h^2 + 6\right)\lh^2+  \frac{1}{2}c_s^2\lhs^2
\eea
\bea
\label{EQ:beta_hs}
16\pi^2\beta_{\lhs}^{(1)} = 4c_hc_s\lhs^2 + 6\left(c_h^2+1\right)\lh\lhs  - \frac{3}{2}\left(3g^2 + g'^2\right)\lhs + 6y_t^2\lhs  + 6c_s^2\ls\lhs
\eea
and \bea
\label{beta_s}
16\pi^2\beta_{\ls}^{(1)}  = \frac{1}{2}(c_h^2 + 3)\lhs^2  +18c_s^2 \ls^2
\eea

The RG equations for the non-minimal coupling can be derived following \cite{Buchbinder:book}, as in \cite{Lerner:2009xg} (see also \cite{Clark:2009dc}).
\be{stuff44} 16\pi^2 \frac{d\xi_s}{dt} =  \left(3+c_h\right)\lhs\left(\xi_h+\frac{1}{6}\right)
+\left(\xi_s+\frac{1}{6}\right)6c_s \ls
\ee
and
\bea 16\pi^2 \frac{d\xi_h}{dt} = \left(\left(6+6c_h\right)\lh + 6y_t^2 - \frac{3}{2}(3g^2 + g'^2)\right)\left(\xi_h+\frac{1}{6}\right)+ \left(\xi_s+\frac{1}{6}\right)c_s\lhs
\eea

The equations for the gauge and Yukawa couplings are
\be{EQ:newg}
\beta_{g} = -\frac{39 - c_h}{12}g^3 + \frac{g^3}{16\pi^2}\left(\frac{3}{2}g'^2 + \frac{35}{6}g^2 + 12g_3^2 - \frac{3}{2}c_h y_t^2\right),
\ee
\bea
\beta_{g'} = \frac{81 + c_h}{12} g'^3 + \frac{g'^3}{16\pi^2}\left(\frac{199g'^2}{18} + \frac{9g^2}{2} + \frac{44g_3^2}{3} - \frac{17c_h y_t^2}{6}\right),
\eea
and
\bea \label{EQ:newyt}
\beta_{y_t} &=& y_t\left(-\frac{9}{4}g^2 - \frac{17}{12}g'^2 - 8g_3^2 + \left(\frac{23}{6} + \frac{2}{3}c_h\right) y_t^2\right) +\frac{y_t}{16\pi^2}\Bigg[ -\frac{23}{4}g^4 - \frac{3}{4}g^2g'^2 \nonumber \\
& &   + \frac{1187}{216}g'^4+ 9g^2g_3^2 + \frac{19}{9}g'^2g_3^2 -108g_3^4 +  \left(\frac{225}{16}g^2 + \frac{131}{16}g'^2 + 36g_3^2\right) c_h y_t^2
\nonumber \\
& &  + 6\left(-2c_h^2 y_t^4 - 2c_h^3y_t^2 \lh + c_h^2\lh^2\right)   \Bigg].
\eea

\subsection*{(iii) Slow roll parameters}

      The Jordan frame RG improved effective potential of \eq{Jpot} is transformed to the Einstein frame in order to compute the slow-roll parameters. Its dependence on $\mu \equiv \phi$ during inflation is then given by the variation of the couplings from RG equations as well as the explicit dependence on the inflaton $\phi$.
We can then relate the slow-roll parameters of the quantum corrected effective potential directly to the RG equations.

Starting with the definition
\be{w5}
\tdeps  = \frac{M_p^2}{2} \left( \frac{1}{U} \frac{dU}{d\chi_{\phi}} \right)^2 =  \frac{M_p^2}{2} \left(\frac{d\phi}{d\chi_\phi}\right)^2 \left( \frac{dU}{d\phi}\frac{1}{U} \right)^2,
\ee
we use
\be{w5c}
\frac{1}{U}\frac{dU}{d\phi} = \frac{4}{\phi\Omega^2} + \frac{1}{\phi}\left( \frac{1}{\lp}\frac{d\lp}{dt} - \frac{4\gamma_{\phi}}{\Omega^2(1+\gamma_{\phi})}- \frac{2\phi^2 G_\phi^2}{\Omega^2M_p^2}\frac{d\xi_{\phi}}{dt}\right)
\ee
to give
\be{EQ:epsWIL}
\tdeps  = \frac{M_p^2}{2\phi^2} \left(  \frac{d\phi}{d\chi_\phi}\right)^2 \left(\frac{4}{\Omega^2} + \frac{1}{\lp}\frac{d\lp}{dt} - \frac{4\gamma_{\phi}}{\Omega^2(1+\gamma_{\phi})}- \frac{2\phi^2 G_\phi^2}{\Omega^2M_p^2}\frac{d\xi_{\phi}}{dt}\right)^2.
\ee
It should be noted that $\gamma_s = 0$ and $G_s = 1$. We have also used $\frac{dG_h}{dt} = -\frac{G_h(t)\gamma_{H}(t)}{1+\gamma_{H}(t)}.$

To find $\tdeta$, we start with the definition
\be{w10}
\tdeta = \frac{M_p^2}{U}\frac{d^2U}{d\chi_\phi^2} = \frac{M_p^2}{U}\left(\frac{d\phi}{d\chi}\right)^2\frac{d^2U}{d\phi^2} + \frac{M_p^2}{U}\frac{d^2\phi}{d\chi^2}\frac{dU}{d\phi}  
\ee
As this expression will be complicated, we assume $c_\phi = 0$ is a good approximation during inflation (when we evaluate $n$ and $r$), since $c_\phi \simeq 0$ from \eq{EQ:supressionfac}. Then we have
\bea
\frac{1}{U}\frac{d^2U}{d\phi^2} &= &-\frac{1}{\phi}\frac{1}{U}\frac{dU}{d\phi} + \frac{1}{U^2}\left(\frac{dU}{d\phi}\right)^2 - \frac{8\xi_\phi G^2}{\Omega^4M_p^2} - \frac{8G^2}{(1+\gamma_h)\Omega^4M_p^2}\frac{d\xi_\phi}{dt}\nonumber \\
& &  + \frac{2\phi^2 G^4}{\Omega^4M_p^4}\left(\frac{d\xi_\phi}{dt}\right)^2 - \frac{2G^2}{\Omega^2M_p^2}\frac{d^2\xi_\phi}{dt^2} +  \frac{8\xi G^2(2\gamma_h+\gamma_h^2)}{\Omega^4 M_p^2 (1+\gamma_h)^2} \nonumber \\
& & + \frac{1}{\phi^2}\left(-\frac{1}{\lh^2}\left(\frac{d\lh}{dt}\right)^2 + \frac{1}{\lh}\frac{d^2\lh}{dt^2} - \frac{4}{\Omega^2(1+\gamma_h)^2}\frac{d\gamma_h}{dt}  \right)
\eea
where
\bea
\frac{d^2\lh}{dt^2} &\simeq& - \frac{1}{1+\gamma_h}\frac{d\gamma_h}{dt}\frac{d \lh}{dt} + \frac{1}{16\pi^2(1+\gamma_h)}\left[12\lh\frac{d\lh}{dt} - 24y_t^3\frac{dy_t}{dt} + \frac{9}{2}g^3\frac{dg}{dt} \right. \nonumber \\
 & & + \frac{3}{2}g'^3\frac{dg'}{dt} + \frac{3}{2}gg'^2\frac{dg}{dt} + \frac{3}{2}g^2g'\frac{dg'}{dt}+ \left(-9g^2-3g'^2+12y_t^2\right)\frac{d\lh}{dt} \nonumber \\
 & & \left.+ \left(-18g\frac{dg}{dt} - 6g'\frac{dg'}{dt} + 24y_t\frac{dy_t}{dt}\right)\lh  \right], 
\eea
\bea
\label{EQ:dgammadt}
\frac{d\gamma_h}{dt} & \simeq & -\frac{1}{16\pi^2}\left(\frac{9g}{2}\frac{dg}{dt} + \frac{3g'}{2}\frac{dg'}{dt} - 6y_t\frac{dy_t}{dt}\right) - \frac{1}{(16\pi^2)^2}\left[\frac{271}{8}g^3\frac{dg}{dt} - \frac{9}{8}gg'^2\frac{dg}{dt} \right.\nonumber \\
& &   - \frac{9}{8}g^2g'\frac{dg'}{dt}  - \frac{5}{2} \left(\frac{9}{2}g\frac{dg}{dt} + \frac{17}{6}g'\frac{dg'}{dt} + 16g_3\frac{dg_3}{dt}\right)y_t^2 \nonumber \\
& & \left. - 5\left(\frac{9}{4}g^2 + \frac{17}{12}g'^2 + 8g_3^2\right)y_t\frac{dy_t}{dt}\right]
\eea
and
\be{w8}
\frac{d^2\xi_\phi}{dt^2} \simeq \frac{1}{\xi_\phi + 1/6}\left(\frac{d\xi_\phi}{dt}\right)^2 + \left(\frac{\xi_\phi + 1/6}{16\pi^2}\right)\left(6\frac{d\lh}{dt} + 12y_t\frac{dy_t}{dt} - 9g\frac{dg}{dt} - 3g'\frac{dg'}{dt}\right).  
\ee
In these expressions,
\be{w12}
\frac{d^2\phi}{d\chi^2} = \frac{1}{\Omega^3}\left(\frac{d\phi}{d\chi}\right)^4 \frac{d\Omega}{d\phi} \left[1  + 3M_p^2 \left(4\left(\frac{d\Omega}{d\phi}\right)^2 -  \frac{d^2(\Omega^2)}{d\phi^2}\right)\right]
\ee
where
\be{imp5}
\frac{d\Omega}{d\phi} = \frac{1}{2\Omega} \frac{\xi\phi G^2}{M_p^2}\left( 2 - \frac{2\gamma}{1+\gamma} + \frac{1}{\xi}\frac{d\xi}{dt}\right)
\ee
and
\be{w14}
\frac{d^2(\Omega^2)}{d\phi^2}  = \frac{2\Omega}{\phi}\frac{d(\Omega)}{d\phi} \frac{1+3\gamma_h}{1+\gamma_h} - \frac{\xi G^2}{M_p^2} \left(\frac{2}{(1+\gamma_h)^2} \frac{d\gamma_h}{dt} + \frac{1}{\xi^2} \left(\frac{d\xi}{dt}\right)^2 - \frac{1}{\xi} \frac{d^2 \xi}{dt^2}\right).
\ee

\section*{Appendix B : Reheating in $S$-inflation}

\renewcommand{\theequation}{B-\arabic{equation}}
 \setcounter{equation}{0}

     Reheating in $S$-inflation is similar to reheating in Higgs Inflation. In both models reheating occurs primarily through preheating via stochastic resonance of the inflaton to bosons, which subsequently annihilate to relativistic Standard Model fields.
In the case of Higgs Inflation this occurs via excitation of gauge boson modes by the oscillating Higgs inflaton, $\chi_{h}$, in which case the coupling is the (known) weak gauge coupling $g$. In the case of $S$-inflation the oscillating $\chi_{s}$ field
excites the Higgs bosons via the (unknown) coupling $\lhs$. In the analysis of \cite{Bezrukov:2008ut}, the gauge boson modes are approximated by scalar modes. This analysis may therefore be directly applied to the case of Higgs boson modes excited by the oscillating $\chi_{s}$ field.
It is also possible for the inflaton to excite its own modes via the self coupling ($\lh$ and $\ls$ for Higgs Inflation and $S$-inflation respectively, where $\lh$ is known and $\ls$ is a free parameter). In the following we will adapt the
method discussed in the appendix of \cite{Bezrukov:2008ut} to the case of $S$-inflation, discussing only the key results and the differences.

    In $S$-inflation, Higgs bosons are created when $\chi_{s}$ passes through zero at times $t_{j}$. Between $t_{j-1}$ and $t_{j}$ the
occupation number of mode $k$ remains constant at $n_k^j$. This changes to
$n_k^{j+1}$ after $\chi_{s}$ passes through zero at $t_{j}$,  where \cite{Bezrukov:2008ut}
\be{evo4c}
n_k^{j+1} = \frac{|R_k|^2}{|D_k|^2} + \frac{1+|R_k|^2}{|D_k|^2} n_k^j + 2 \sqrt{1+n_k^j}\sqrt{n_k^j}\frac{|R_k|}{|D_k|^2}\cos(\theta^j_{tot})
\ee
Here $R_{k}$ and $D_{k}$ are `reflection' and `transmission' coefficients relating coefficients $\alpha_k^j$ and $\beta_k^j$ of the Higgs mode function $\psi_{k} \equiv a^{3/2} \phi_{k}$ before and after $\chi_{s}$ passes through zero at $t_{j}$, and
$\theta_{tot}^j = -2\theta^j_k - 2 \left(\frac{2}{3}\kappa^3 + \frac{\pi}{4}\right) +\arg \alpha_k^j - \arg \beta_k^j$, where $\theta_{k}^j$ is the phase of the mode function. There are two regimes, depending on the size of $n_k$.

\noindent (i) $n_k \ll 1$: During the time $\Delta t = \frac{\pi}{\omega}$ between zero crossings, $n_k^{j} \simeq \frac{|R_k|^2}{|D_k|^2}$ scalars $\phi_i$ are produced. (Here $\phi_{i}$ ($i = $ 1,...4) are the four real scalars in the Higgs doublet.) The majority of these particles decay (to relativistic Standard Model particles) before the next zero crossing and there is no significant transfer of energy to radiation.

\noindent (ii) $n_k \gg 1$: it is now the second and third terms of \eq{evo4c} that are important and we can write (for each $\phi_i$)
\be{ii1}
n_k^{j+1} = e^{2\pi \mu_k} n_k^j
\ee
where $\mu_k$ is an average over the angle $\theta^j_{tot}$, which is assumed to be completely randomly distributed. This can be calculated, giving
\bea
\label{solveint}
\mu_k
& = & \frac{1}{2\pi}\left[\ln\left(\frac{|R_k| + 1}{|R_k|}\right) + i - \ln\left(|D_k|^2\right) \right] \nonumber \\
& \simeq & -\frac{1}{2\pi}\ln\left(|D_k|^2\right)~.
\eea
In order to have a resonance, the reflection coefficient, $|R_k|$, must be reasonably large. This means that $|D_k|$ is small (from the requirement $|R_k|^2 + |D_k|^2 = 1$) and this gives the final line of \eq{solveint}. We are interested in the rate of change of the total number of particles produced, so we must integrate over $k$. Thus, as the change in $n_k$ during the time between zero crossings of the inflation, $\Delta t = \frac{\pi}{\omega}$, is $\Delta n_k \simeq 2\pi \mu_k n_k$, the total rate of production of Higgs scalars is
\bea
\frac{dn}{dt} \simeq \frac{\omega}{\pi}\int_0^\infty\left(2\pi \mu_k n_k \right)\frac{d^3 k}{(2\pi)^3} ~,
\eea
where $n$ is the total number of $\phi_{i}$ scalars.
This integral is not calculated exactly,
instead it is assumed that $\mu_k$ is well approximated by $\mu(k= 0) \equiv B$ \cite{Bezrukov:2008ut}, which is found to have the value $B = 0.045$.
Summing all four real Higgs bosons in $H$ gives for the rate of change of total number density of Higgs bosons $n_T$
\be{ii2}
\frac{dn_T}{dt} \sim 4\times2\omega B n = 2\omega B n_T.
\ee
In \cite{Bezrukov:2008ut}, the components of the $W^{\pm}$ and $Z$ bosons produced via preheating by the Higgs boson are treated as scalars. This is not strictly true for that case but it is correct for our calculation of the production of scalar Higgs bosons via preheating.

The Higgs bosons which are produced at the zero crossings can decay in the adiabatic regime between zero crossings where their mass is large. The Higgs has a large mass because the background $\chi_{s}$ field is large. Therefore it can decay via the Yukawa coupling to fermion pairs, primarily the top quark. The decay width of the Higgs bosons (with mass averaged over an $\chi_{s}$ oscillation) is given by \cite{Donoghue:bk}
\be{evo6}
\Gamma_{\phi_{i}} = \frac{N_c g^2 m_f^2}{32\pi m_W^2} \langle m_\phi^2 \rangle^{1/2} \equiv C \langle m_\phi^2 \rangle^{1/2} \simeq \frac{1}{17.65}\langle m_\phi^2 \rangle^{1/2}
\ee
where $\langle m_\phi^2 \rangle = \langle\frac{\lhs s^2}{2}\rangle$. The Higgs bosons are non-relativistic when produced but their decay and annihilation products are relativistic.

Denoting the amplitude of the oscillation of $\chi$ by $X$, we can investigate whether reheating can complete before quartic oscillations begin, defined to be $X_{CR} = \sqrt{\frac{2}{3}} \frac{M_p}{\xi_s}$. Once the decay of the produced Higgs bosons becomes subdominant to their production, an exponential regime can begin, corresponding to Bose-enhanced generation of Higgs bosons. Up to this point, no significant energy transfer from $\chi_{s}$ to the Higgs bosons occurs since Higgs decay prevents a large Higgs occupation number. Reheating completes fairly rapidly after this, as once the Higgs bosons are produced through stochastic resonance, they can annihilate and produce a thermal background, draining most of the energy from the $\chi_{s}$ oscillations and causing its amplitude $X$ to rapidly decrease to $X_{CR}$ at which point the quartic oscillations begin. The upper limit on the time of reheating is the time at which the decay and the exponential production are equal. Using \eq{evo6} and \eq{ii2} we find that the two processes are equal at
\be{re1}
X_{sto} \simeq \frac{2\pi B^2}{C^2} \frac{\ls}{\lhs} X_{CR} \simeq 4.0 \frac{\ls}{\lhs}\;X_{CR}~.
\ee
This is the absolute maximum value of $X$ at which reheating occurs via production of $H$ bosons. After $\chi_{s} = X_{sto}$, $\chi_{s}$ will rapidly decrease to $X_{CR}$, therefore $\rho_{rad} > \rho_{inf}$ at $X_{CR}$. The calculation assumes instant annihilation of the $H$-bosons to relativistic particles and a very efficient stochastic resonance, therefore $X_{sto}$ is an upper limit of the value of $X$ at which reheating occurs. For this reason all estimates of $T_R$ from stochastic resonance are {\em upper limits}. It is interesting to note that increasing the Higgs-$s$ coupling, $\lhs$, causes reheating to occur {\em later}. This is because a large $\lhs$ means that $m^2_h$ is large, so the Higgs bosons can decay for a longer period, preventing the exponential regime from beginning and therefore delaying reheating.

The time of reheating is dependent on the ratio of couplings $\frac{\ls}{\lhs}$. Requiring this process of reheating to occur before $X = X_{CR}$ gives
\be{re1b}
\ls > 0.25\lhs~.
\ee
If this condition is not fulfilled, reheating is not ruled out. It could either occur by directly producing excitations of the inflaton, as discussed below, or possibly during the quartic potential regime. Since our goal is to show that reheating in $S$-inflation can occur for natural values of
the couplings, we will conservatively assume that \eq{re1b} is satisfied. If it is not satisfied and if reheating can occur in the quartic regime, a somewhat lower temperature of radiation domination $T_R$ would be expected.

An alternative method of draining energy from the background field is the direct production of excitations of the inflaton $\chi_{s}$, which is possible as the inflaton potential changes at small $\chi_{s}$ to $V \propto \chi_{s}^4$. The mode equation can be solved perturbatively, assuming the number of particles produced is small. The $s$ particles produced are relativistic, with energy density \cite{Bezrukov:2008ut}
\be{ed1}
\rho_{excitation} = \frac{3}{11}\frac{\omega^5}{2\pi^3} t ~.
\ee
From this, if the production of excitations is the only process of reheating then radiation domination occurs when the $\chi_{s}$ oscillation amplitude is $X = X_{ex}$, where
\be{ed2}
X_{ex} \simeq 7\sqrt{\ls}X_{CR}.
\ee
If this mechanism of reheating is to occur before the inflaton potential becomes quartic, we require
\be{ed3}
\ls > 0.019.
\ee
The relativistic $\chi_{s}$ particles are expected to subsequently annihilate to Higgs bosons, so producing the thermal background of SM particles. This is most efficient for large $\lhs$. Both of these limits are shown in \fig{couplim}.

\end{document}